\documentclass[
  journal=pasa, %largetwo,
  manuscript=research-article,
  year=2025,
  volume=XX,
]{cup-journal}

\usepackage{amsmath}
\usepackage{commath}
\usepackage{upgreek}
\usepackage[nopatch]{microtype}
\usepackage{booktabs}
\usepackage{multirow}
\usepackage{longtable}
\usepackage{supertabular}
\usepackage{arydshln}
\usepackage{siunitx}
\usepackage{subcaption}
\usepackage{float}
\usepackage{rotating}

\usepackage[colorlinks=true]{hyperref}
\newcommand*{\fullref}[1]{\hyperref[{#1}]{\nameref*{#1}}}
\hypersetup{
    colorlinks=true,
    linkcolor=purple,    
    urlcolor=.,
    citecolor=orange,
    }

\newcommand{\fig}{Fig.}
\newcommand{\figs}{Figs.}

\newcommand{\sect}{Section}
\newcommand{\sects}{Sections}
\newcommand{\Sect}{Section}

\newcommand{\tab}{Table}

\newcommand{\eqn}{equation}

\newcommand{\mJypbm}{\text{mJy}\,\text{beam}$^{-1}$}
\newcommand{\uJypbm}{$\mu$\,\text{Jy beam}$^{-1}$} 
\newcommand{\Jypbm}{\text{Jy beam}$^{-1}$} 

\title{GaLactic and Extragalactic All-sky Murchison Widefield Array survey eXtended (GLEAM-X) III: Galactic Plane}

\author{S. Mantovanini}
\affiliation{International Centre for Radio Astronomy Research, Curtin University, Bentley WA 6102, Australia}
\email[S. Mantovanini]{silvia.mantovanini@postgrad.curtin.edu.au}

\author{N. Hurley-Walker}
\affiliation{International Centre for Radio Astronomy Research, Curtin University, Bentley WA 6102, Australia}

\author{K. Ross}
\affiliation{Australian SKA Regional Centre (AusSRC), Curtin University, Bentley WA 6102, Australia}
\alsoaffiliation{International Centre for Radio Astronomy Research, Curtin University, Bentley WA 6102, Australia}

\author{S. W. Duchesne}
\affiliation{CSIRO Space and Astronomy, PO Box 1130, Bentley WA 6102, Australia}

\author{G. Anderson}
\affiliation{International Centre for Radio Astronomy Research, Curtin University, Bentley WA 6102, Australia}

\author{T. J. Galvin}
\affiliation{CSIRO Space and Astronomy, PO Box 1130, Bentley WA 6102, Australia}

\affiliation[A]{International Centre for Radio Astronomy Research, Curtin University, Bentley WA 6102, Australia}
\affiliation[B]{Australian SKA Regional Centre (AusSRC), Curtin University, Bentley WA 6102, Australia}
\affiliation[C]{CSIRO Space and Astronomy, PO Box 1130, Bentley WA 6102, Australia}


\begin{document}

\begin{abstract} 
We present the third data release for the Galactic and Extragalactic All-Sky Murchison Widefield Array eXtended (GLEAM-X) survey, covering $\approx 3800$~deg$^2$ of the southern Galactic Plane (GP) with $\ang{233} < l < \ang{44}$ and $|b| < \ang{11}$ across a frequency range of $72 - 231$\,MHz divided into 20 sub-bands. GLEAM-X observations were taken using the ``extended'' Phase-\textsc{ii} configuration of the Murchison Widefield Array (MWA), which features baselines ranging from approximately $12$~m to $5$~km. This configuration limits sensitivity to the diffuse structure of the GP, with an angular resolution range of about $45^{''}$ to $2^{'}$. To achieve lower noise levels while being sensitive to a wide range of spatial scales ($45^{''} - \ang{15}$), we combined these observations with the previous Galactic and Extragalactic All-Sky Murchison Widefield Array (GLEAM) survey. For the area covered, we provide images spanning the whole frequency range. A wide-band image over $170-231$\,MHz, with RMS noise of $\approx 3-6$~\mJypbm and source position accuracy within 1 arcseconds, is then used to perform source-finding, which yields 98,207 elements measured across $20 \times 7.68$\,MHz frequency bands. The catalogue is 90$\%$ complete at 50\,mJy within $\ang{233} < l < \ang{324}$ and at 125\,mJy in $\ang{290} < l < \ang{44}$, while it is $99.3\%$ reliable overall. All the images and the catalogue are available online for download. 
\end{abstract}

\section{Introduction} 
Radio astronomy surveys along the Galactic Plane (GP) have undergone significant upgrades over the last century, advancing our understanding of the Milky Way thanks to sky coverage, sensitivity, and resolution improvements. %The first radio surveys covering the GP were carried out in the 1950s at approximately 400\,MHz with low angular resolution from $10^{\circ}$ to $20^{\circ}$ \citep{Wielebinski2004}. 
GP surveys are essential in understanding the complex structure of the Milky Way, its composition, how it evolves and what emission processes come into place. Spectral index studies are key to differentiating the emission mechanisms occurring in various objects \citep{Cavallaro2017} and understanding how they evolve and interact. Mid-to-high radio frequency bands can provide insights into the thermal emission of star-forming regions \citep{Brunthaler2021}. In contrast, low radio frequencies are particularly useful for negative spectrum sources (assumed to be $S \propto \nu^{\alpha}$ with $\alpha$ having negative values), such as pulsars \citep{Swainston2021}, or to analyse free-free absorption and synchrotron self-absorption traces along the plane, useful in constraining the distribution of Cosmic Ray (CR) electrons and Galactic Magnetic Field (GMF) \citep{Su2018,Nord2006}. Moreover, supernova remnants (SNRs) offer key insights into the life cycle and evolution of stars in our Galaxy \citep{Dubner2015}.

Mid-frequency surveys, such as the ongoing Evolutionary Map of the Universe \citep[EMU;][]{Norris2021} at 944\,MHz using the Australian Square Kilometre Array Pathfinder \citep[ASKAP;][]{Hotan2021} telescope, and the new SARAO~\footnote{South African Radio Astronomy Observatory} MeerKAT Galactic Plane Survey \citep[SMGPS;][]{Goedhart2024} at 1300\,MHz, demonstrate the impressive results of surveys with modern wideband mid-frequency receivers in terms of resolution and sensitivities, obtaining a few arcseconds and $10-20$~\uJypbm, respectively \citep{Anderson2025}. 

Despite the notable success of the latest surveys, we still face certain limitations. While these surveys often aim for full sky coverage, the small field of view at high frequencies (GHz regime) makes the data acquisition process less efficient, requiring more observing time to cover large areas. Another limitation is the maximum angular scale that mid-radio surveys are sensitive to, which is intrinsically tied to the size of the interferometer dishes and their array configuration, which limits the baseline coverage for the largest spatial scales. 
%EMU concentrates on a single frequency range to produce a complete southern sky map at gigahertz frequencies. Similarly, SMGPS presents limited sky coverage. Furthermore, although it has a wider frequency range (from 500\,MHz to 2\,GHz), neither covers lower frequencies (below a few hundred megahertz) at which absorption effects are measurable (for Galactic temperatures, densities, and magnetic field strengths). 
\cite{Goedhart2024} analysed the problem for the SMGPS in detail, highlighting that the spectra for sources with a radius greater than $2'$ can be significantly inaccurate and should not be used. The MeerKAT frequency-dependent sensitivity recovers a higher fraction of the flux densities for extended sources at the lowest frequency. The result is an apparent steepening in the spectral index for non-point-like sources. Although the Parkes Galactic All-Sky Survey (PEGASUS) project showed the first results of addressing this limitation by providing images sensitive to larger scales, which can be combined with EMU data to produce more accurate sky images, this project is still ongoing and will require time to be completed. As a result, neither of these surveys is sensitive to a broad range of spatial scales crucial for science cases, such as detecting and performing spectral analysis of diffuse structures within the Galaxy, like star-forming regions \citep{Medina2024}, supernova remnants and Galactic bubbles \citep{Ingallinera2014}.

On the other hand, low-frequency surveys provide critical insights into the non-thermal sky. A key strength is offered by low-frequency interferometers, such as the established Giant Metrewave Radio Telescope \citep[GMRT][]{Swarup1991}, which offers moderate angular resolution with sensitivity to both compact and moderately extended emission (up to $\approx \ang{1}$ at 150\,MHz), and latest instruments such as the LOw Frequency ARray (LOFAR) and the Murchison Widefield Array \citep[MWA,][]{Tingay2013,Wayth2018}, which employ dense core configurations with predominantly short baselines to enhance sensitivity to large angular scales. This feature allows low-frequency surveys to recover extended emission from diffuse structures, such as SNRs and \textsc{hii} regions \citep{Castelletti2007}, far more effectively than higher-frequency surveys like EMU or SMGPS. 

However, it is worth mentioning that ionospheric effects become increasingly problematic at these lower frequencies, introducing strong, frequency-dependent phase distortions ($\propto \nu^{-2}$) that vary across the field of view. These distortions can lead to position shifts and smearing of sources in the images if not properly corrected \citep{Intema2009}. The severity of these effects increases during periods of high ionospheric activity and toward lower elevations. LOFAR is actively implementing complex, direction-dependent calibration schemes to correct spatially varying ionospheric delays \citep{vanWeeren2016}, essential for achieving accurate astrometry.

MWA, the low-frequency precursor of the SKAO, presents an extensive field-of-view ($10^2-10^3$\,sq.deg.) and a wide frequency band $80-300 \, \text{MHz}$, ideal for all-sky surveys south of declination (Dec) $+30^{\circ}$. The instrument Phase~\textsc{i} configuration (with a baseline range varying approximately from 7~m to 2.5~km - see Appendix~\ref{app:uvcov}) enabled the creation of the Galactic and Extragalactic All-sky MWA \citep[GLEAM; ][]{Wayth2015} survey, which, thanks to the abundance of short baselines, has spatial scale sensitivity to ($2' - 15^{\circ}$) reaching noise levels of 10~\mJypbm in the extragalactic sky \citep{Hurley2017} and 50 -- 100~\mJypbm along the GP \citep{Hurley2019c}. GLEAM-GP enabled detailed studies of synchrotron emission, thermal absorption, and spectral turnovers, providing new insights into the structure and conditions of the interstellar medium. Its sensitivity to large-scale, diffuse emission has allowed the mapping of the GMF \citep{Polderman2020}, investigations of cosmic ray propagation \citep{Su2018}, and the discovery of previously unidentified SNRs and \textsc{hii} regions \citep{Hurley2019b,Anderson2017}. 

The ``extended'' Phase~\textsc{ii} configuration of MWA (baseline range from $\approx 12$~m to 5~km - see Appendix~\ref{app:uvcov}), on the other hand, used doubled length baselines to collect the data for the GLEAM-eXtended \citep[GLEAM-X; ][]{Hurley2022,Ross2024} survey, which is incrementally being made available to the community as portions are completed. The longer baselines allow GLEAM-X to effectively capture smaller scales ($45^{''} - 20'$), and its lowered confusion limit enables better sensitivity over long integrations ($\sim 1$~\mJypbm). 

Here we present the GLEAM-X Galactic Plane survey (GLEAM-X-GP), which combines data from the GLEAM and GLEAM-X surveys, achieving a synergistic combination of resolution and sensitivity to an ample range of spatial scales across an observing band of 72 -- 231\,MHz. Several approaches have been developed to combine data from multiple surveys to address the problem of missing short spacings in interferometric imaging, e.g. feathering in the image plane \citep{Cotton2017}. Some large-area surveys, such as the GLObal view on STAR formation in the Milky Way \citep[GLOSTAR][]{Brunthaler2021} and VLA Galactic Plane Survey \citep[VGPS][]{Stil2006}, have implemented short-spacing corrections at multiple frequencies to improve the recovery of extended emission. In this work, we apply joint deconvolution to offer a consistent reconstruction of structures, avoiding accentuating deconvolution artefacts or calibration errors that might be present in individual image methodologies. In doing so, we will have multi-frequency sky coverage of the southern plane with sensitivity to a broad range of spatial scales, enabling accurate flux density measurements even for larger sources. %The dataset is suitable for conducting spectral studies at extremely low frequencies (below 300\,MHz) on specific astronomical objects. We will mainly focus on detecting and analysing extended objects, including SNRs and \textsc{hii} regions.

This paper presents the data reduction used to produce images covering the GLEAM-X-GP and an associated catalogue over the longitude range of $\ang{233} < l < \ang{44}$ and within latitudes $\abs{b} < \ang{11}$. \sect~\ref{sec:observations} briefly summarises the observations. \sect~\ref{sec:reduction} describes the calibration, imaging and mosaicking procedures. \sect~\ref{sec:catalogue} describes the derivation of a source catalogue along with a thorough assessment of its quality. \sect~\ref{sec:gp} discusses the science applications that this release can achieve. \Sect~\ref{sec:conclusion} concludes with a short summary.

\section{Observations}\label{sec:observations}
GLEAM and GLEAM-X observations were collected using similar strategies to efficiently cover the entire sky south of $\ang{+30}$ in Dec. The observations were performed in drift-scan mode, in which the telescope pointing remained fixed while the sky drifted through the instrument's primary beam. For telescopes like the MWA, pointing refers to the electronically formed beams created by applying time delays or phase shifts to the signals from individual components to direct the telescope's sensitivity toward a desired region of the sky without moving the instrument. Seven pointings have been used at Decs centred on $\ang{-72}$, $\ang{-55}$, $\ang{-40.5}$, $\ang{-26.7}$, $\ang{-13}$, $\ang{+1.6}$, and $\ang{+18.3}$. The frequency range is divided into five bands $30.72$\,MHz wide: 72--103\,MHz, 103--134\,MHz, 139--170\,MHz, 170--200\,MHz, and 200--231\,MHz, avoiding the band around 137\,MHz due to contamination by the OrbComm satellite constellation. Each pointing was observed for approximately two minutes per band, iterating over the whole frequency range every 10~minutes. Observations were repeated over multiple nights; during each night, bright calibrator sources were observed repeatedly over the five frequency channels to obtain amplitude and phase gain solutions \citep[listed by][]{Hurley2017}. 

The ``extended'' Phase~\textsc{ii} configuration of the MWA \cite[described by][]{Wayth2018} resulted in a smaller synthesised beam, which reduced the confusion limit in GLEAM-X from $\approx 2$\,mJy\,beam$^{-1}$ to $\approx 0.3$\,mJy\,beam$^{-1}$ at 200\,MHz (Paterson et al., submitted). These advancements enabled longer integration times, allowing deeper observations and lower noise levels \citep{Hurley2022}. GLEAM observations were conducted over $\approx 28$ nights from August 2013 to June 2014, with additional observations in 2015 to re-observe regions where ionospheric conditions were particularly poor. GLEAM-X was observed over a three-year period from January 2018 to October 2020 for a total of 113 nights. \tab~\ref{tab:obsids} provides the first four lines of the table summarising all the observations used to create the final image of the GP. The full table can be downloaded from AAO Data Central.

\section{Data reduction} \label{sec:reduction}
GLEAM and GLEAM-X data have been jointly deconvolved, and all visibilities are used together to create one (dirty) image for deconvolution to achieve sensitivity to all the angular scales between $45^{''} - 15^{\circ}$ while increasing the $S/N$ of the sources. We employed the Image Domain Gridding \citep[IDG;][]{vanderTol2018} algorithm implemented in \textsc{WSClean} \citep[$w$-stacking clean;][]{Offringa2014} for the data combination, efficiently converting extensive datasets from the visibility to the image domain. When combining data from multiple observations, IDG uses direction and time-dependent convolution kernels during the gridding stage to align the contributions from each observation onto a common grid. This approach ensures that direction-dependent (ionospheric) distortions are minimised. This is advantageous when compared to standard tools (e.g. TCLEAN in \textsc{CASA} \cite{Casa2022} and \textsc{WSClean}), which only apply these corrections in a post-imaging phase or do not support them at all. Furthermore, IDG enables a \textit{hybrid} mode with Graphical Processing Units (GPUs) to perform the gridding and fast Fourier transforms, significantly decreasing the computational costs and making joint imaging feasible for large datasets. 
%IDG incorporates ionospheric distortions during the gridding stage, improving consistency in the final image, especially in multi-observation data where angular separations and time-dependent variations can be large.

\begin{figure*}[hbt!]
    \centering
    \includegraphics[width=1\linewidth]{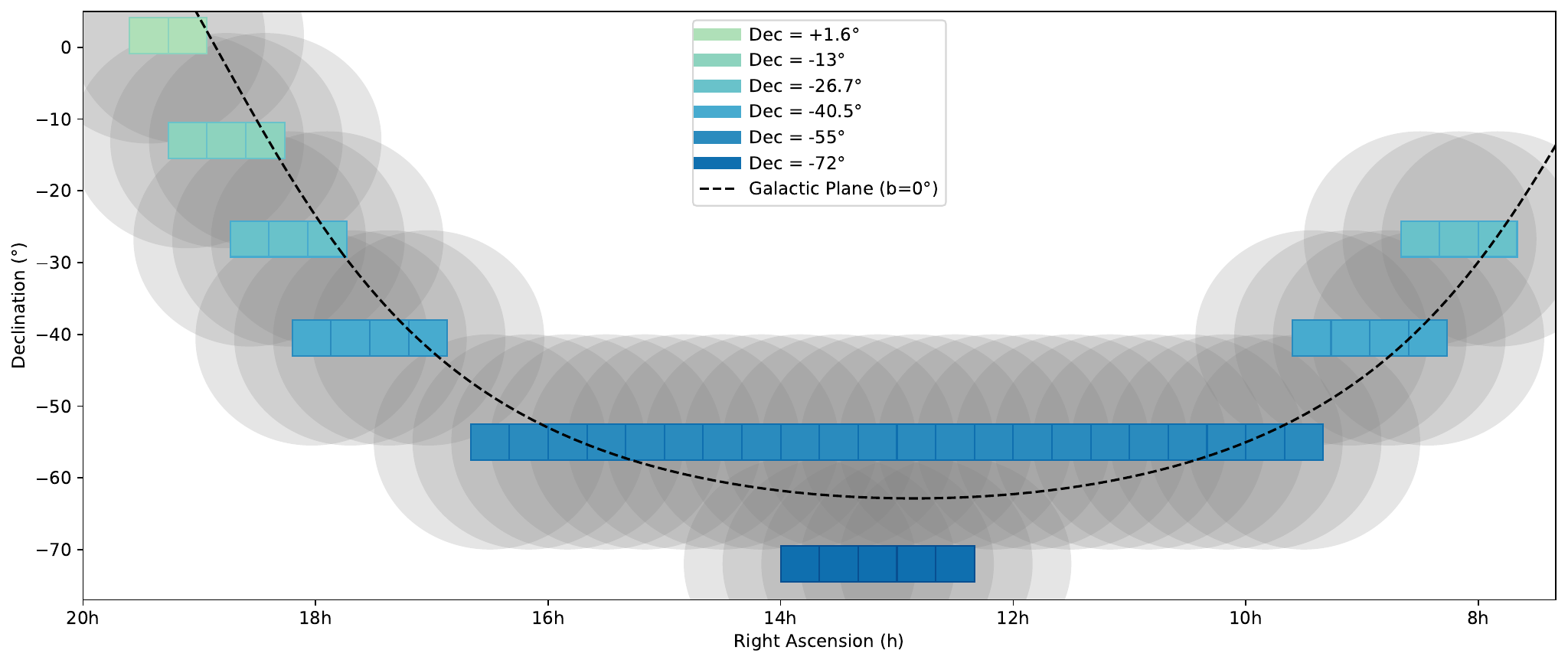}
    \caption{Representation of the grouping strategy for GLEAM and GLEAM-X. Each rectangle defines a group of 10--15 observations whose pointing centres fall within the coordinates of the boxed region, ensuring complete coverage over the GP across the declination range. Circular regions indicate individual fields of view, with overlaps between adjacent circles providing continuous coverage. The dashed black line marks the GP ($b = \ang{0}$) in FK5 coordinates.}\label{fig:grouping}
\end{figure*}

We divided the GLEAM and GLEAM-X observations into small groups comprising approximately 10--15 observations, all with pointing centres with the same Dec and spanning a RA range of $\approx \ang{5}$, as represented in \fig~\ref{fig:grouping}. While each group is ideally balanced to include data from both surveys to ensure adequate representation, this was not always feasible due to the quality constraints (see \sects~\ref{sub:calib} to \ref{sub:astrometric}) of the observations.

\begin{figure*}[hbt!]
    \centering
    \includegraphics[width=1\linewidth]{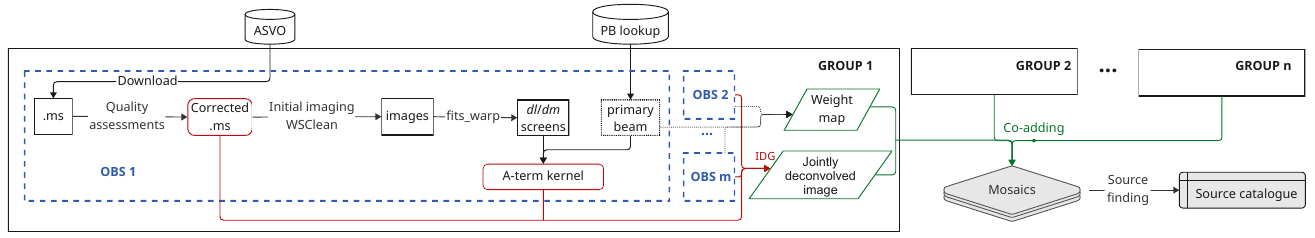}
    \caption{Flow diagram of the data reduction procedure. The diagram shows the main processing steps, from downloading the observations of each group to adding all the IDG group images to form the final mosaic.}\label{fig:pipeline}
\end{figure*}

The data reduction performed in this work is heavily based on the GLEAM-X pipeline~\footnote{https://github.com/GLEAM-X/GLEAM-X-pipeline} described in full by \cite{Hurley2022}, and summarised in the flowchart of \fig~\ref{fig:pipeline}. Some improvements to data processing have been introduced to increase the image quality of the GP pointings and account for the presence of bright and extended radio sources that dominate the plane. Furthermore, we implemented additional steps to prepare the data to be jointly deconvolved and then mosaicked to form the final image. Here, we summarise the reduction steps, highlighting whether they have been changed from the previous methodology.

\subsection{Calibration and flagging}\label{sub:calib}
We began assessing the quality of the observations covering the plane. We calibrated each 2-minute observation via transferring amplitude and phase gains solutions obtained by the closest bright calibrator during the same night. An in-field calibration using the GLEAM sky model is executed if the calibrator observation gains are unavailable; for this purpose, we used all the baselines except the shortest ($<100$\,m) and longest ($> 2.5$\,km) to, respectively, avoid contamination from diffuse emission, and to match the resolution of the GLEAM-based calibration sky model \citep[described by ][]{Hurley2022}. We visually inspected the solutions for every observation to assess the calibration quality; we discarded all observations with anomalies or unrealistic values of amplitude and phase gains over frequency. 

To account for Radio Frequency Interference (RFI) contamination, we implemented an extra flagging stage for those observations in the lowest frequency channel (72 -- 103\,MHz), where more RFI is observed. Medians are calculated for Stokes~\textsc{I} amplitudes of the visibilities for each measurement set using a window of 55~channels at a time. Every channel with a median value greater than 0.5~$\sigma$ - where $\sigma$ is the standard deviation within the window - is flagged (removed). We additionally flagged two fine channels at the edges of each coarse channel (24, each 1.28\,MHz wide) to avoid leakage. This ensures that only high-quality data is used for calibration and imaging.

Another origin of contamination is the presence of bright radio sources that fall in the sidelobes of our observation primary beam. We used the approach described by \cite{Ross2024} to deal with this issue. For each observation, we phase-rotate the visibilities using \textsc{chgcentre} to match the coordinates of the source, then use \textsc{WSClean} to form a shallow image $128 \times 128$ pixels in size to create a model of the object. Once the model is created, it is subtracted from the observation measurement set visibilities, which are then phase-rotated back to their original location. Every observation goes through this process, and all the bright sources above the horizon and falling in their sidelobes are subtracted. If the distance of the source from the pointing centre is $\leq \ang{30}$, the process does not apply as the source will be properly cleaned during the imaging stage. Centaurus~A is dealt with with a slightly different approach because of its high surface brightness and diffuse extension. For this reason, Centaurus~A is subtracted from the visibilities even if it is on the edge of the observations. The subtraction is performed for observations with a pointing within $\ang{170} < RA < \ang{250}$ to reduce the artefacts and the root-mean-square (RMS) noise level of the final image. In this case, the size of the shallow box is increased to $600 \times 600$ pixels to better model Centaurus~A's large-scale emission.

\subsection{Initial snapshot imaging}\label{sub:imaging}
Initial 2-minute images of each observation of the GP are generated by \textsc{WSClean} using the settings described in the survey description papers (\cite{Hurley2019a} and \cite{Hurley2022} for GLEAM and GLEAM-X observations, respectively). We do not directly use these images for the final mosaic; they are utilised to assess the quality of the observation data by inspecting potential artefacts caused by residual RFI or bright sources and calibration performance. The detected position of the sources is saved as a reference point for ionospheric corrections, which is essential for enabling the joint deconvolution of each group of observations. These corrections ensure accurate source localisation, as described in \sect~\ref{sub:astrometric}. In the following, we will briefly list the parameters that vary between the two surveys and the parameters we have implemented in this work:

\begin{itemize}
    \item \textbf{\textsc{-size}}: GLEAM pixel images of $4000 \times 4000$ and $8000 \times 8000$ pixel images for GLEAM-X to encompass the field-of-view down to 10\% of the primary beam and a frequency-dependent pixel scale to ensure each image has 3.5 -- 5 pixels per FWHM of the restoring beam;
    \item \textbf{\textsc{-weight}}: different Briggs robust parameters \citep{Briggs1995} as a good compromise between resolution and sensitivity for the two configurations of the MWA: $-1$ (close to uniform weighting) is chosen for GLEAM to maximise the resolution and suppress the PSF sidelobes, while $+0.5$ is chosen for GLEAM-X to optimise the overall RMS noise - i.e to improve sensitivity, which is ideal for detecting fainter signals;
    \item \textbf{\textsc{-multiscale-scale-bias}}: increased the multiscale bias parameter \citep{Offringa2017} from the default value of 0.6 to 0.7 only for GLEAM-X observations to reduce the effects of diffuse emissions, which sometimes otherwise led to convergence problems or poorly cleaned point sources. 
\end{itemize}

To avoid affecting the remaining observations of each group, we discarded the 2-minute observations that show an exceptionally high RMS or residual artefacts we could not remove in the previous steps (see \sect~\ref{sub:calib}). Common problems are the presence of strikes across the image: vertical or horizontal are most likely caused by residual RFI; radial and originating from bright sources in the field caused by deconvolution errors; bright streaks radiating from outside the field of view caused by sources in the edges of the primary beam or the sidelobes; and regular striping around bright sources, or distortion of sources caused by poor calibration.

\subsection{Astrometric calibration} \label{sub:astrometric}
To measure the position shifts induced by the ionosphere, we applied a similar methodology described in \cite{Hurley2019a} and later in \cite{Hurley2022}. We used \textsc{aegean} \citep{Hancock2012,Hancock2018} on each 2-minute observation to determine the position of point sources in the image. We cross-matched it with a reference catalogue composed of a sparse (no internal matches within $3'$) and unresolved (integrated to peak flux density ratio of < 1.2) sample of the NRAO VLA Sky Survey \citep[NVSS,][]{Condon1998} at 1.4\,GHz and the Sydney University Molonglo Sky Survey \citep[SUMSS,][]{Mauch2003} at 843\,MHz respectively for the northern and southern sky. \textsc{fits \textunderscore warp} \citep{Hurley2018} is ideal for accurately creating a model of the sources' pixel offsets by a cross-match between the snapshot and the reference catalogue. However, NVSS and SUMSS do not cover the sky area between $\ang{252} < l < \ang{353}$ and $b < |\ang{11}|$; we selected the Rapid ASKAP Continuum Survey \citep[RACS;][]{Mcconnell2020,Hale2021} to fill in the gap in this region. For this purpose, we applied the criteria mentioned above to the RACS catalogue and selected a subset of only bright sources with integrated flux above 50\,mJy at 888\,MHz. We concatenated the catalogues, excluding internal matches of $3'$. The position shifts along both axes in the sky plane (\textit{l} and \textit{m}) are saved into FITS-format cubes that will be applied to the final IDG image to correct for the ionospheric effects as will be described in \sect~\ref{sec:joint}. 

We analysed the distribution of pipeline failures across the five 30\,MHz bands. The results indicate lower frequencies (72--103\,MHz) exhibit higher failure rates, particularly during the imaging stage, whereas higher frequencies (170--200\,MHz) show fewer losses overall. Among the pipeline stages, calibration and post-imaging checks account for most failures, highlighting how these steps are critical data loss points in the pipeline. In appendix~\ref{app:iono}, we report the percentage of observations discarded due to high ionospheric activity as a function of time (see \fig~\ref{fig:iono}). After assessing the quality of the observations, $57\%$ were retained for the final imaging, while the remaining $43\%$ were discarded due to quality issues. Of the successful observations, $53\%$ came from GLEAM-X and $47\%$ from GLEAM, reflecting the balanced contribution of both surveys to the final dataset.

\subsection{Joint deconvolution}\label{sec:joint}
We now run the IDG algorithm implemented in \textsc{WSClean} on GLEAM and GLEAM-X observation groups. The algorithm requires a configuration file characterising the direction-dependent effects we must account for in the cleaning process (the $a$-term corrections). The file points to the FITS cubes generated in \sect~\ref{sub:astrometric} to correct the position of sources in the image by applying a position shift based on the previously saved information. During cleaning, we also correct for the MWA primary beam \citep[Full Embedded Element model][]{Sokolowski2017}, an instrumental effect of the antennae, which are assumed to have identical responses over the whole array.
%Without considering this issue, sources will attenuate in flux, appearing fainter.
Following the previous data releases \citep[][]{Hurley2017,Ross2024}, the primary beams are calculated and interpolated as required using code developed by \cite{Morgan2021} for each element of the group. Finally, the beam correction is averaged across the full set of observations, improving the final image quality. 
%available from github~\footnote{https://github.com/johnsmorgan/mwa$\_$pb$\_$lookup}

IDG also accounts for $w$-terms in wide-field imaging caused by the large field of view (can extend up to $\ang{30}$ depending on the frequency band) by incorporating $w$-corrections directly into the gridding process. Without these corrections, the standard two-dimensional Fourier transform of visibilities assumption (i.e., flat-sky approximation) becomes invalid, leading to artefacts around sources far from the phase centre, showing smearing. Ultimately, for each observation, we performed a visibilities phase rotation using \textsc{chgcentre} to match the average pointing centre of all the observations in the group.

In the following, we list the parameters that vary compared to the 2-minute observation \textsc{WSClean} imaging run (presented in \sect~\ref{sub:imaging}) and the parameters added to implement the new algorithm:

\begin{itemize}
    \item  \textbf{\textsc{-auto-mask}}: initial masking threshold to control which regions are cleaned, set to 3, meaning only pixels with brightness above three times the estimated noise level are included during deconvolution. Cleaning will continue until the \textbf{\textsc{-auto-threshold}} stopping criterion of every major deconvolution iteration is reached. A threshold factor of 0.5 is chosen to reduce losses in flux densities for extended objects in the GP. 
    %\textbf{\textsc{-auto-threshold}}: threshold factor for stopping cleaning of 0.5 to reduce the losses in flux densities for the extended objects in the GP. This stopping criterion is directly related to the residual noise level, as the cleaning won't stop until the threshold factor times the standard deviation of the residual image before the start of every major deconvolution iteration is reached. It is used in combination with \textbf{\textsc{-auto-mask}} of 3, which sets the initial masking threshold to control which regions are cleaned;
    \item \textbf{\textsc{-size}}: pixel images of $8000 \times 8000$ and a frequency-dependent pixel scale (0.5/central frequency expressed in MHz) to ensure each image has 3.5 -- 5 pixels per FWHM of the restoring beam;
    \item \textbf{\textsc{-weight}}: a Briggs robust parameter \citep{Briggs1995} of $-0.25$ as a good compromise between resolution and sensitivity for the two configurations of the MWA;
    \item \textbf{\textsc{-idg-mode}}: IDG is used in a hybrid mode using 1 NVIDIA Tesla V100 Tensor Core GPU per job, which increased processing efficiency.
\end{itemize}

\subsection{Mosaicking} \label{sub:mosaic}
%45 x 5 channels x 4 subbands
The same basic image quality checks applied in Section~\ref{sub:astrometric} are performed on each of the resulting $~900 \times 7.68$\,MHz IDG group images. We then generated a weight map ($w_i$) for every $i$th IDG group image as:

\begin{equation}
    w_i = \dfrac{\sum \limits_n \dfrac{1}{2}(B_{xx} + B_{yy})}{n},
\end{equation}

which corresponds to the average of the total intensity beam, calculated as the sum of the intensity components along the x-axis $B_{xx}$ and y-axis $B_{yy}$, across all $n$ observations in each group. 

We then used \textsc{swarp} \citep{Bertin2002} to co-add the images, creating an extended mosaic for every $20 \times 7.68$\,MHz frequency channel. Furthermore, we also formed a wide-band 60\,MHz mosaic over 170 -- 231\,MHz for optimal sensitivity and resolution for the source-finding process, as described in Section~\ref{sec:catalogue}; we created additional $5 \times 30.72$\,MHz mosaic images to enhance the $S/N$ and increase the detectability of fainter sources, thereby enabling more accurate estimates of their spectra. For each of the five bands, we then provide the corresponding four 7.68\,MHz sub-band mosaics and one 30.72\,MHz mosaic, along with an additional wideband image combining the top two frequency bands (170--231\,MHz). To ensure a smooth co-addition, we down-weighted the edges of the weight maps to avoid hard cuts in the mosaics by using a declination-dependent sigmoid function ($1/(1 + \exp^{-(\mathrm{Dec}-\mathrm{Dec}_\mathrm{ref})})$). 

All mosaics have been corrected for the blurring effect that can arise due to residual ionospheric distortions, following the method of \cite{Hurley2017}. We first selected compact sources from the reference catalogue (introduced in \sect~\ref{sub:astrometric}) with $S/N > 10$. We performed an initial source-finding on the mosaics using \textsc{aegean} with a seedclip of $10\sigma$. We then cross-matched the reference sample to the output catalogue form \textsc{aegean}; the size and shape of these sources are then measured from the GP mosaics. We used these sources to determine the PSF map. The process involved interpolating over the sources utilising a Healpix projection (corresponding to pixels roughly $\ang{3}$ across) and averaging each pixel with its neighbours. A smooth map of the major and minor axes, position angle and ``blur'' factor is obtained, where the latter corresponds to the ratio between the measured PSF volume and the expected PSF volume from projection effects alone. We multiply the final mosaic by this (position-dependent) factor to normalise the flux density scales so that the peak and integrated flux densities are consistent. Over Galactic latitudes $\abs{b} < \ang{4}$, there are insufficient point-like sources, so we perform an interpolation of the PSF components ($a_{PSF}$, $b_{PSF}$, PA) and the blur factor.
%to mitigate the effect using the mosaic PSF map previously interpolated

\begin{figure}[t]
    \centering
    \includegraphics[width=1\linewidth]{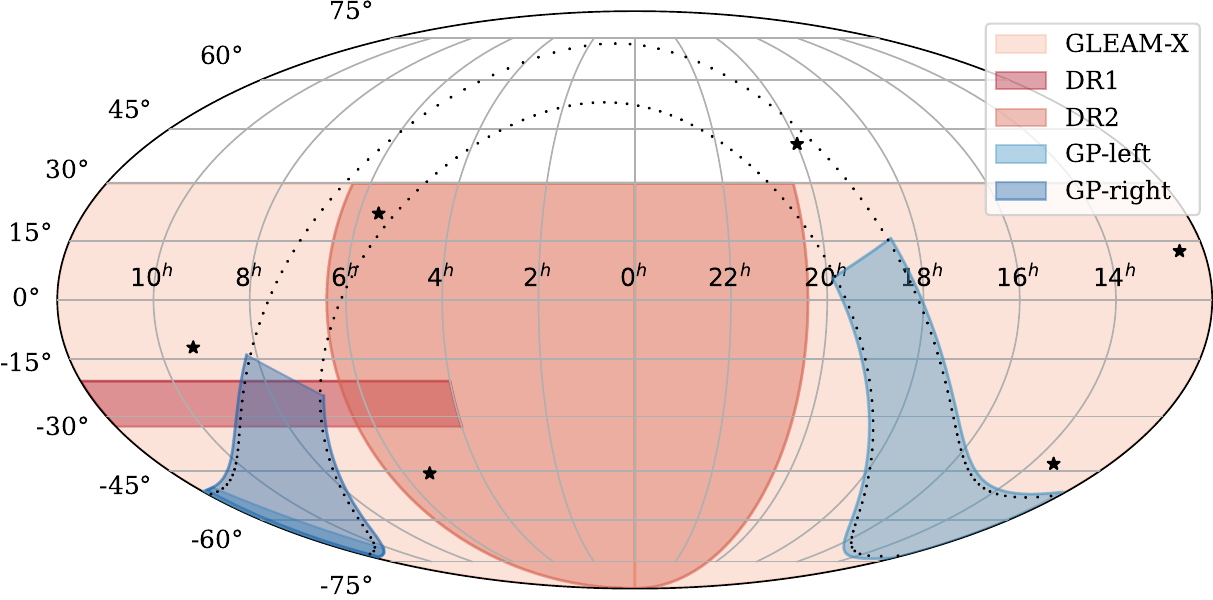}
    \caption{Sky coverage of the GLEAM-X survey, with the light orange region indicating the total area observed. The dark red region represents the area covered in the first data release \citep[DR1;][]{Hurley2022}, and the light red region highlights the area covered by the second data release \citep[DR2;][]{Ross2024}. In contrast, in light and dark blue are reported the area covering $\ang{290} < l < \ang{44}$ and $\ang{233} < l < \ang{324}$ respectively, presented in this data release (GP-left, GP-right). The black stars indicate the location of bright radio sources listed from left to right as they appear on the map: Hydra~A, Crab, Pic~A, Cygnus~A, Centaurus~A, and Virgo~A. The black dotted line shows the GP within $|b|<\ang{10}.$}\label{fig:coverage}
\end{figure}

In order to minimise distortions caused by projection effects, the mosaics have been generated in two sky regions covering $\ang{290} < l < \ang{44}$ and $\ang{233} < l < \ang{324}$, within latitudes of $\abs{b} < \ang{11}$, using Zenithal Equal Area (ZEA) projection. This maintained a consistent, equatorial-coordinate PSF, which simplifies the formation of a source catalogue \citep{Calabretta2002}. The regions of the sky covered in this data release and in the previous two GLEAM-X releases are presented in \fig~\ref{fig:coverage}, where the sky coverage of the whole GLEAM-X survey is highlighted in light orange. In addition, the two regions have been mosaicked together using Galactic coordinates for a complete view of the diffuse structures of the Galactic plane.
%(305 right, 233 left (GC)

\begin{sidewaysfigure*}
    \centering
    \includegraphics[width=1\linewidth]{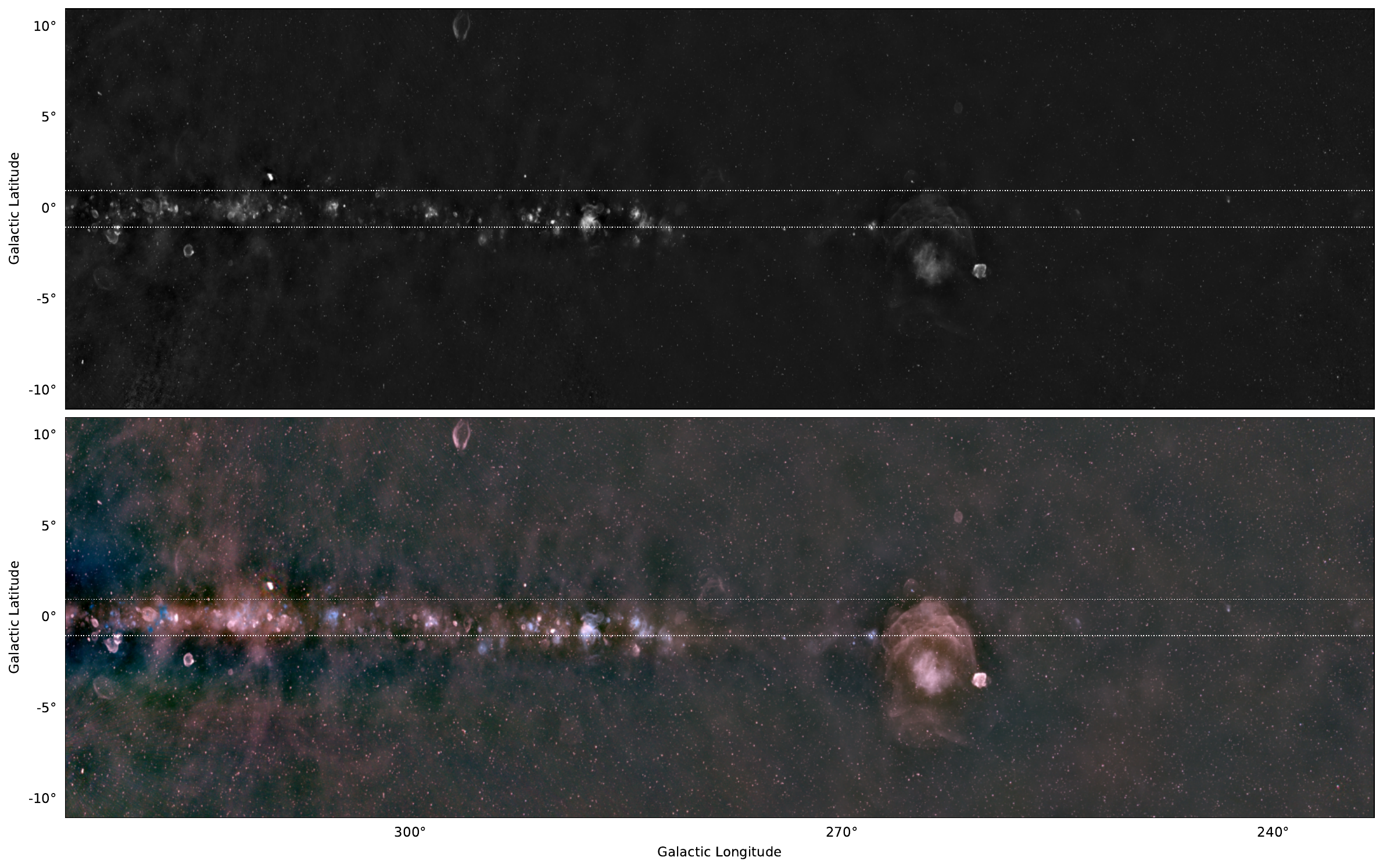}
    \caption{The wide-bandwidth images from the data described in this paper; this illustrative figure shows the  $\ang{233} < l < \ang{324}$ region. The top panel shows the 170--231\,MHz. The bottom panel shows an RGB cube formed of the 72--103\,MHz (R), 103--134\,MHz (G), and 139--170\,MHz (B) data. Dotted white lines indicate $|b| < \ang{1}$. The colour ranges used are -0.025--1.0~\Jypbm and -0.095--1.25~\Jypbm with an arcsinh stretch for the two panels, respectively.}\label{fig:plane-l}
\end{sidewaysfigure*}

\begin{sidewaysfigure*}
    \centering
    \includegraphics[width=1\linewidth]{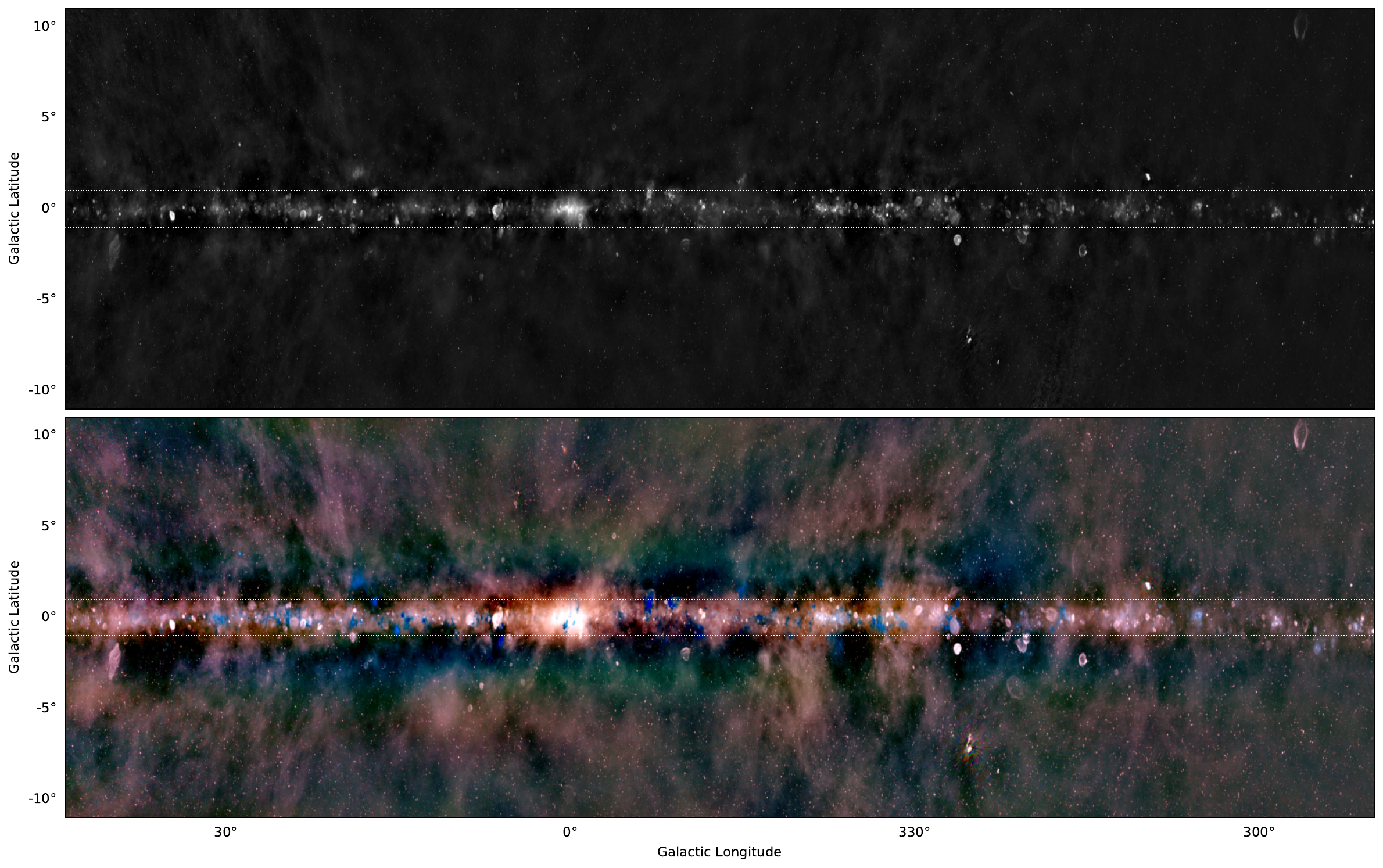}
    \caption{The wide-bandwidth images from the data described in this paper; this illustrative figure shows the  $\ang{290} < l < \ang{44}$ region. The top panel shows the 170--231\,MHz. The bottom panel shows an RGB cube formed of the 72--103\,MHz (R), 103--134\,MHz (G), and 139--170\,MHz (B) data. Dotted white lines indicate $|b| < \ang{1}$. The colour ranges used are -0.025--1.0~\Jypbm and -0.095--1.25~\Jypbm with an arcsinh stretch for the two panels, respectively.}\label{fig:plane-r}
\end{sidewaysfigure*}

%Size= (11+11)*(127+44)=3762
The $26 \times 2$ mosaics generated in \sect~\ref{sub:mosaic} comprise a total of 1,898 2-minute observations across the two regions. Overall, the sky area covered spans $\ang{233} < l < \ang{44}$ and latitudes of $\abs{b} < \ang{11}$, as represented in \fig~\ref{fig:plane-l} and \ref{fig:plane-r}. We limited the surveyed sky primarily driven by two factors. First, the longitude range accounts for observational constraints, particularly the elevation limits in the northern hemisphere and the presence of Cygnus~A, which affects the deconvolution and calibration quality. Sensitivity also decreases at low elevations, making further extensions less effective. We limited the latitude range to $|b| < \ang{11}$ because the joint deconvolution approach is relatively complex, and the extragalactic sky is more easily processed using the basic pipeline (\cite{Ross2024}, Ross et al., in prep). The total area of the data released here is $\approx 3800$~deg$^2$. 

%\begin{figure}[t]
%    \centering
%    \includegraphics[width=1\linewidth]{Figures/Observations.pdf}
%    \caption{Distribution of the pipeline failures across the five frequency channels we employed. Each colour corresponds to a specific processing stage, with failures accumulating from earlier to later stages.}\label{fig:failure}
%\end{figure}

\subsection{Noise levels}
We now analyse the noise properties of the wide-band image to verify that it is Gaussian, a necessary condition for accurate source characterisation. We follow the procedure of \cite{Hurley2019c,Ross2024}, applying it separately to the two regions of the GP: $\ang{290} < l < \ang{44}$ and $\ang{233} < l < \ang{324}$. For each region, we used a representative 16\,deg$^2$ cutout with a small mean RMS noise of $6$~\mJypbm and $3$~\mJypbm, respectively for $\ang{290} < l < \ang{44}$ and $\ang{233} < l < \ang{324}$, and typical source distribution. The 16\,deg$^2$ cutouts are located outside the plane to avoid contamination from bright diffuse structures and are centred at coordinates of $l,b = (\ang{16.4}, \ang{-6.6})$ degrees and $l,b = (\ang{301.6}, \ang{-7.9})$ degrees. 

\begin{figure*}
    \centering
    \begin{subfigure}[b]{\linewidth}
    \includegraphics[width=1\linewidth]{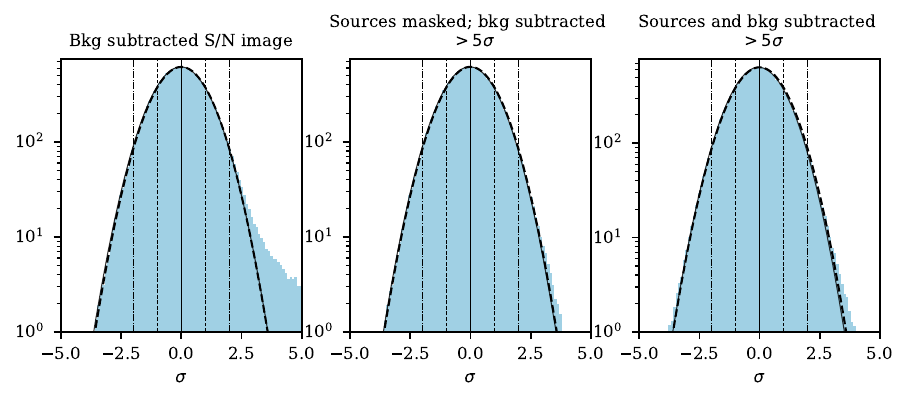}
    \end{subfigure} \\
    \begin{subfigure}[b]{\linewidth}
    \includegraphics[width=1\linewidth]{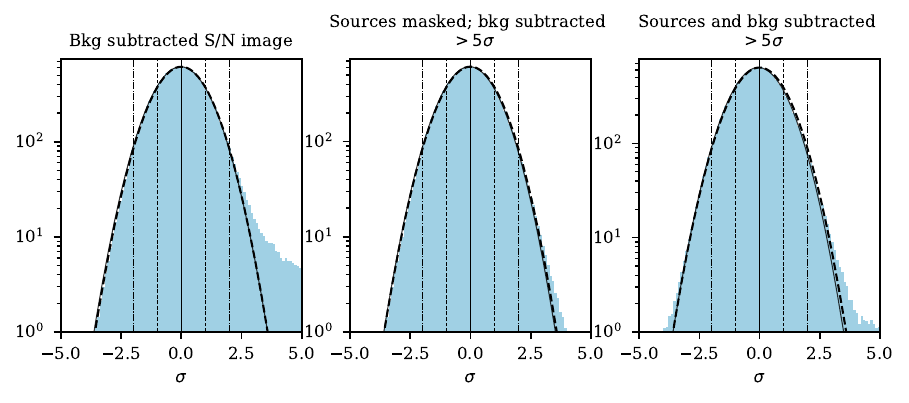}
    \end{subfigure}
    \caption{Pixel distribution for a 16\,deg$^2$ region of the wide-band source finding image covering $170-231$\,MHz. The top row correspond to the region centred in $l,b = (\ang{16.4}, \ang{-6.6})$ for $\ang{290} < l < \ang{44}$, while the bottom row corresponds to the region centred in $l,b = (\ang{301.6}, \ang{-7.9})$ for $\ang{233} < l < \ang{324}$. The RMS noise levels, determined using \textsc{bane}, are $6$~\mJypbm and $3$~\mJypbm, respectively. Each figure consists of three panels: the leftmost panel shows the distribution of pixel $S/N$ values after subtracting the background and dividing by the RMS noise map. The central and right panels show the same distribution but are limited to sources $> 5\sigma$ with faint sources masked using \textsc{aeres} (central panel) or subtracted (right panel). The black solid line represents a Gaussian distribution with $\sigma = 1$ (as measured by \textsc{bane}, while the black dashed line is a fitted Gaussian to the pixel distribution. Vertical solid lines indicate the mean values; dashed lines correspond to $|S/N| = 1\sigma$; and dash-dotted lines correspond to $|S/N| = 2\sigma$.}\label{fig:noise}
\end{figure*}

We extract the cutout images from the mosaic at $170--231$\,MHz, generating background and RMS maps using \textsc{bane}. We subtract the background from the images and perform source extraction with \textsc{aegean}. The source catalogue derived from these images is subsequently subtracted or masked using \textsc{aeres} from the background-subtracted images. In \fig~\ref{fig:noise}, we show the histogram of pixel distribution for the background-subtracted images and the same distribution with faint sources masked and source-subtracted for both the regions of our analyses. The histogram visually assesses the effectiveness of the subtractions applied.

The noise values we obtain at the low latitudes are 5--8 times lower than the GLEAM data published by \cite{Hurley2019c}, thanks to the higher resolution of the GLEAM-X survey provided by the longer baselines of the MWA Phase-\textsc{ii} extended configuration, and the $\sim3\times$ longer total integration time processed in this work. This combination ensures that we do not hit the classical confusion limit. RMS noise levels as a function of longitude across the frequency coverage are shown in the Appendix~\ref{app:noise} for additional insight into the image quality.

\section{Catalogue}\label{sec:catalogue}
We used the same technique described in \cite{Hurley2019a,Hurley2022} to construct a source catalogue; we briefly summarise the process's main steps. We first measured the background and RMS maps of the wide-band image using \textsc{bane}. The grid size in the \textsc{bane} call has been doubled from the default value to account for bright extended structures along the GP, particularly around the Galaxy's centre. A higher grid size value reduces the chances of subtracting the signal from the diffuse structures as background by estimating it over a larger area.

\textsc{aegean} is run with a seedclip of $4\sigma$ to find sources in the wide-band image to create a first source list. We applied a global rescaling to match the GLEAM reference catalogue to ensure a consistent flux density scale across the mosaicked images. The source positions from this list are then used to measure sources in the narrow band images (8\,MHz wide) using a technique called ``priorised'' fitting \citep[see ][ for further details on the process]{Hancock2018}, which takes into account the position-dependent PSF for each image. The list is filtered to retain only sources with peak flux densities $\geq 5 \times$ the local RMS. These steps were performed in both regions, where we divided the GP, producing two catalogues with a substantial overlap region. We introduced six groups in common across both regions for mosaic generation, ensuring an overlap of $\approx \ang{34}$ in longitude. This approach was chosen to maintain consistent image quality and to establish a clear boundary between the two mosaics, allowing for easy and reliable source selection from the catalogues. To merge the two catalogues, we then selected a hard boundary at a longitude of 309 degrees: if a source is located at a lower longitude, we consider the source from the $\ang{290} < l < \ang{44}$ source list; otherwise, if it falls to a higher value, we retain the entry from the $\ang{233} < l < \ang{324}$ source list. The final catalogue contains $98207$ radio sources.

\begin{figure*}
    \centering
    \includegraphics[width=1\linewidth]{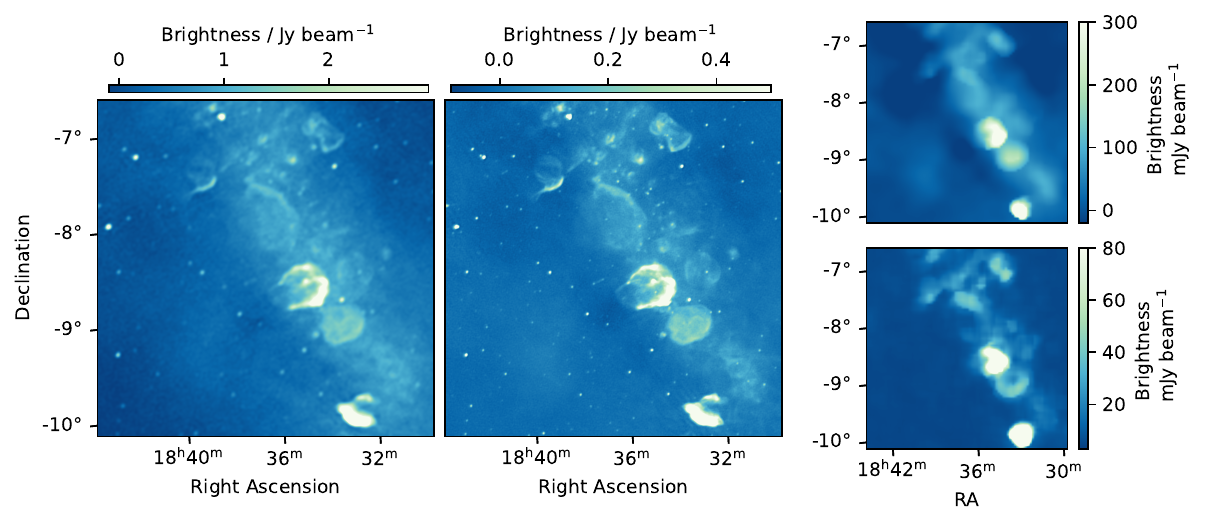}
    \caption{$\approx 16$\,deg$^2$ from the 170--231\,MHz mosaic of GLEAM and this work centred on RA 18h36m and Dec $-\ang{8}$. The region, as seen by GLEAM, is reported in the left panel; the middle shows the wide-band image of this work; and the small panels on the right represent the background (top) and RMS (bottom) maps of the wide-band image. The median RMS is $657$~\mJypbm and $103$~\mJypbm in GLEAM and this work, respectively. \textsc{Aegean} detected $\approx 90$ and $\approx 300$ sources in GLEAM and in this work.
    }\label{fig:bkg}
\end{figure*}

Caution needs to be taken when working with sources within $|b| < \ang{1}$, for which \textsc{bane} does not properly isolate the background levels. \fig~\ref{fig:bkg} illustrates a region of $\approx 16$\,deg$^2$ from the 170--231\,MHz image, along with its background and RMS maps, and the same sky area as observed by GLEAM \citep[][]{Hurley2019a}. Although there is an increase in the number of detected sources compared to the previous survey, rising from $\approx 90$ to $\approx 300$, users are cautioned to carefully examine the flux density values reported in the catalogue and interpret them in context with the images.

%RMS -> median gp 0.007, median gleam 0.03

\subsection{Spectral fitting}\label{sub:fitting}
\begin{figure*}
    \centering
    \begin{subfigure}[b]{0.32\linewidth}
    \includegraphics[width=1\linewidth]{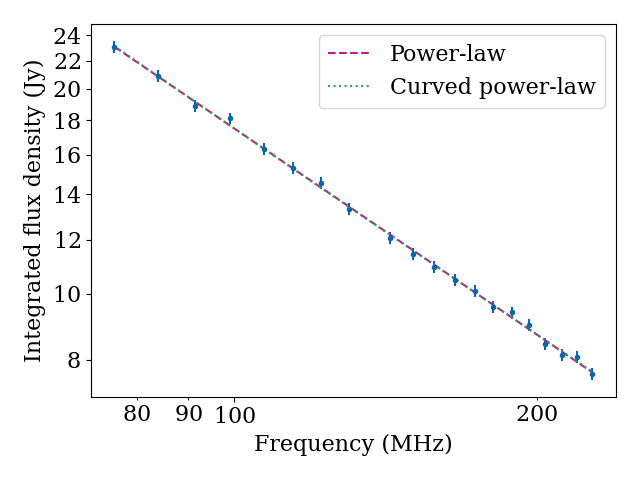}
    \caption{GLEAM-X J175254.4$-$374528}\label{fig:seds:example1}
    \end{subfigure}
    \begin{subfigure}[b]{0.32\linewidth}
    \includegraphics[width=1\linewidth]{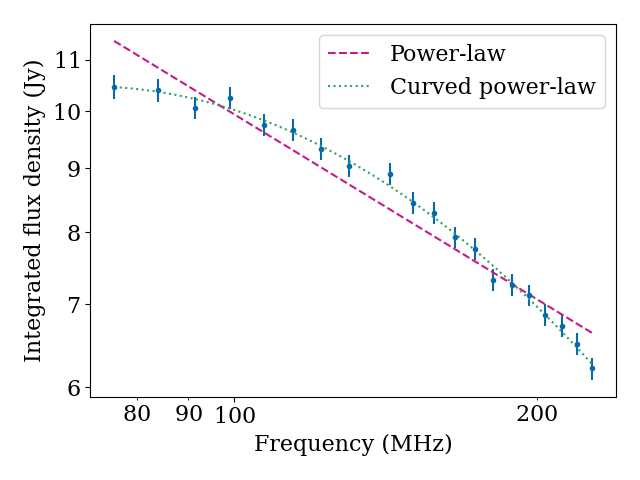}
    \caption{GLEAM-X J165956.9$-$305206.}\label{fig:seds:example2}
    \end{subfigure}
    \begin{subfigure}[b]{0.32\linewidth}
    \includegraphics[width=1\linewidth]{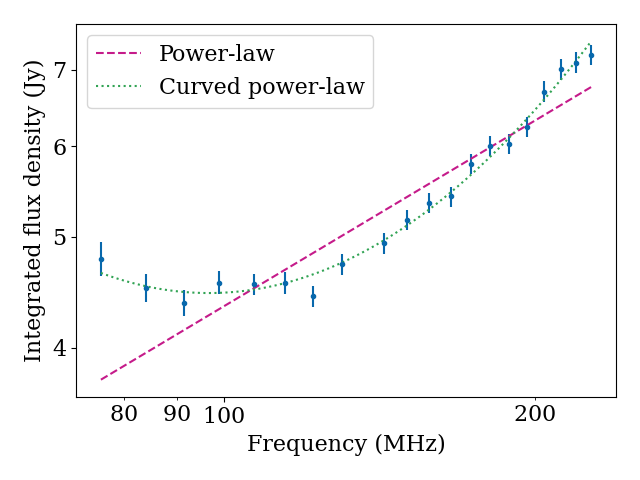}
    \caption{GLEAM-X J183339.9$-$210339.}\label{fig:seds:example3}
    \end{subfigure}
    \caption{Example SEDs for the narrow band measurements of three unresolved point sources. \subref{fig:seds:example1} shows a standard power law spectrum, \subref{fig:seds:example2}--\subref{fig:seds:example3} show convex and concave curved power-law spectra, respectively. The spectral model has been selected by the criteria listed in \sect~\ref{sub:fitting}.
    %All three panels have both axes reported in log scale.
    }\label{fig:seds}
\end{figure*}

We also performed spectral fitting, following the procedure described by \cite{Ross2024}, and report the fit spectral energy distributions (SEDs) parameters of sources from the catalogue. Thanks to the ``priorised'' fitting approach (\sect~\ref{sec:catalogue}), we were able to extract flux densities across the 20 narrow bands. We applied two spectral models: a simple power law parametrised as $S_{\nu} = S_{\nu_{0}} (\nu / \nu_0)^{\alpha}$, and a curved power law, obtained by multiplying $S_{\nu}$ by $\exp(q\,\ln(\nu/\nu_0)^2)$ where $q$ is related to convex or concave curve respectively for positive or negative values. We used a curved power law to estimate the sample's possible peaked-spectrum sources (PSS). The model selection is based on reduced-$\chi^2$ values; if both models fit well, the one with the lower reduced-$\chi^2$ is chosen. A simple power law successfully fitted 89655 sources presenting at least 15 integrated flux density measurements with a $\chi^2$ goodness-of-fit above $99\%$ likelihood confidence; the other 3132 were fitted with a curved spectrum after satisfying the condition of having a reduced-$\chi^2$ lower compared to the reduced-$\chi^2$ for a power law model and having $q\,\Delta q \geq 3$ with $|q| > 0.2$. \fig~\ref{fig:seds} displays example SEDs for each model. The flux densities used in the SEDs and the spectral indices obtained are consistent with those reported in the GLEAM-GP catalogue, as discussed in \sect~\ref{sub:gleam}, validating the flux scale across the full frequency range.

%92787 fitted by a power law (89655) or curved power law (3132)

\begin{figure}
    \centering
    \includegraphics[width=1\linewidth]{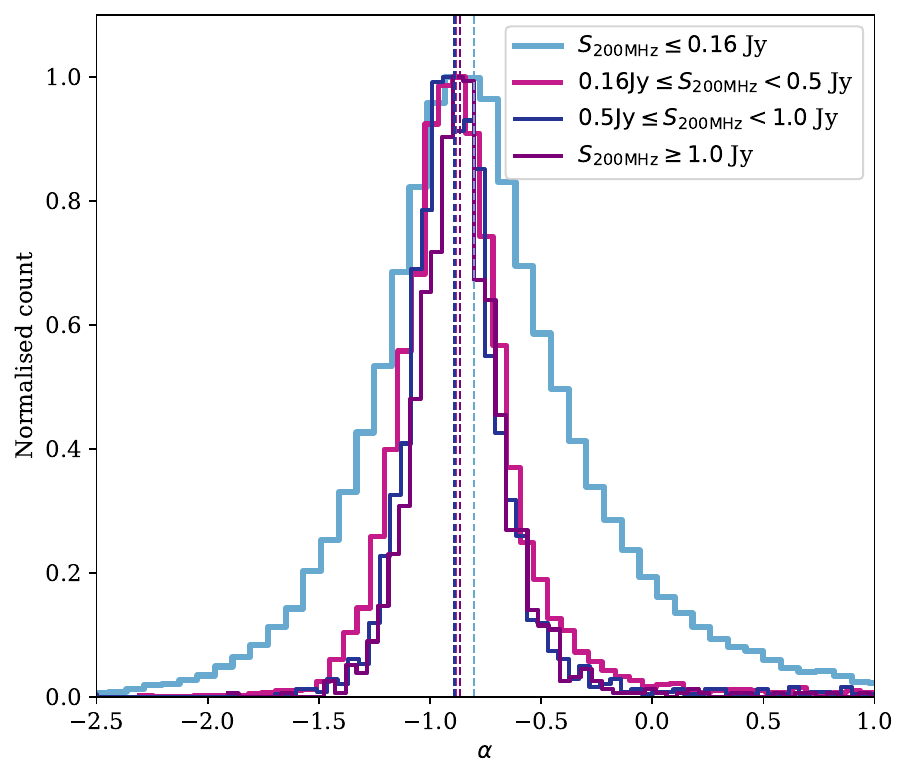}
    \caption{Distribution of the spectral index $\alpha$ as a function of the flux density. The cyan line shows sources with $S_{200\,MHz} < 0.16\,Jy$, the magenta line shows sources with $0.16\,Jy \leq S_{200\,MHz} < 0.5\,Jy$, the blue line shows sources with $0.5\,Jy \leq S_{200\,MHz} < 1.0\,Jy$, and the purple line shows sources with $S_{200\,MHz} \geq 1.0\,Jy$. The dashed lines of the same colour highlight the median value for every flux density bin, and they respectively correspond to -0.80, -0.87, -0.88, and -0.86.}\label{fig:alpha}
\end{figure}

We estimated the median spectral indices for different flux density ranges and reported the results in \fig~\ref{fig:alpha}. Our findings show an excess of steep-spectrum sources along the GP, particularly towards the Galactic centre, with a systematic difference in spectral index between the longitude ranges. By combining data from NVSS at 1400\,MHz and the Tata Institute for Fundamental Research GMRT Sky Survey \citep[TGSS,][]{Intema2017} at 147\,MHz, \cite{Degasperin2018} analysed this trend in detail, suggesting excess steep-spectrum sources originate from undiscovered radio pulsars, probably missed from past searches due to scattering along the line of sight caused by turbulent ionised gas that broadens the observed pulsation width beyond detectability with traditional time-domain surveys.
%-0.71 right and -0.85 left

\subsection{Astrometry}
\begin{figure}
    \centering
    \includegraphics[width=1\linewidth]{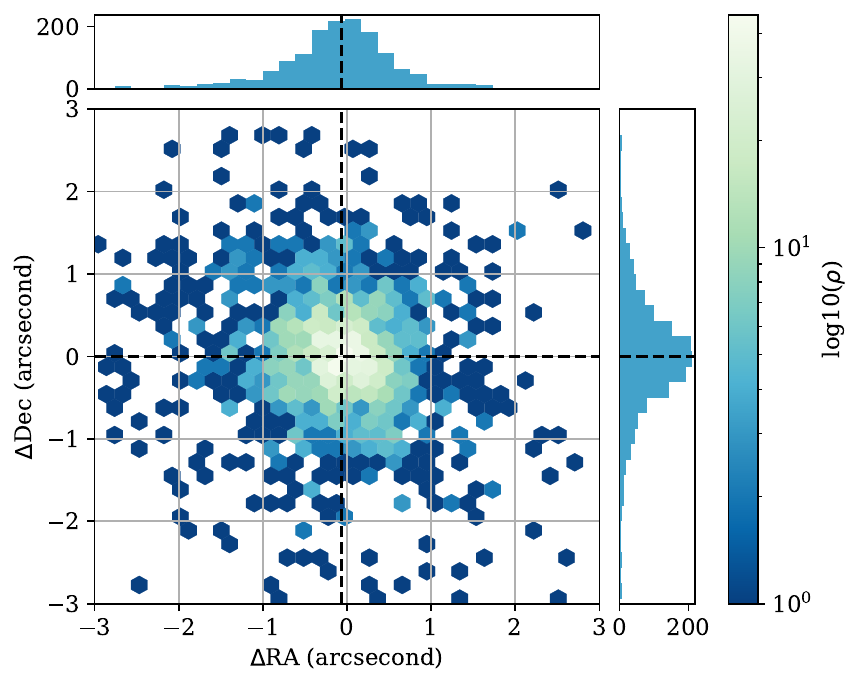}
    \caption{The astrometric offsets of 8421 sources which are isolated, compact, and $> 10\,\sigma$ after they have been cross-matched against the reference catalogue described in \sect~\ref{sub:astrometric}. Black vertical and horizontal lines correspond to the mean offset in the RA and Dec directions and have a value of $-148 \pm 980$\,mas and $+7 \pm 980$\,mas, respectively. Histograms show the counts for the astrometry offset measures in each direction. The colour bar represents the density of points ($\rho$) on a logarithmic scale.}\label{fig:offsets}
\end{figure}

To characterise the precision of the final catalogue, we use the reference list described in \sect~\ref{sub:astrometric} to cross-match with a subset of the GP source catalogue at 200\,MHz and determine the offsets in their position. The subset comprehends sources with a high signal-to-noise ratio ($\geq10$; hereafter referred to as $S/N$), isolated (with no internal matches within $10'$ and unresolved ($(a_{src} \times b_{src}) / (a_{PSF} \times b_{PSF}) < 1.1$). The density distribution of the astrometric offsets is shown in \fig~\ref{fig:offsets}. The average RA and Dec astrometric offsets, with respect to the reference list, are $-148 \pm 980$\,mas and $+7 \pm 980$\,mas, respectively.
%These averages have been calculated considering the source's position in the reference catalogue as an exact value.
%As mentioned in \sect~\ref{sub:astrometric}, we corrected the position of the sources in the individual 2-minute observation images based on the same reference catalogue, biasing the results we report here.

\subsection{Reliability and completeness}
To assess the reliability of our catalogue, we performed the same analysis as \cite{Hurley2022} and \cite{Ross2024}. The procedure checks for false detections and was conducted separately for the $\ang{290} < l < \ang{44}$ and $\ang{233} < l < \ang{324}$ regions. We limit the reliability and the following completeness simulations to $4^\circ < |b| < 11^\circ$  as we do not expect them to be accurate in the presence of the bright diffuse structures of the GP.

\begin{figure*}
    \centering
    \begin{subfigure}[b]{0.48\linewidth}
    \includegraphics[width=1\linewidth]{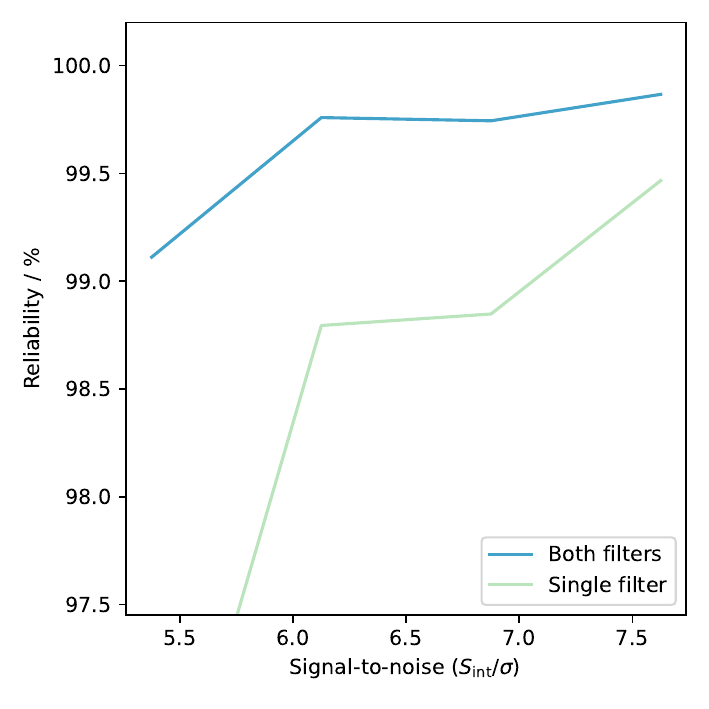}
    \end{subfigure}
    \begin{subfigure}[b]{0.48\linewidth}
    \includegraphics[width=1\linewidth]{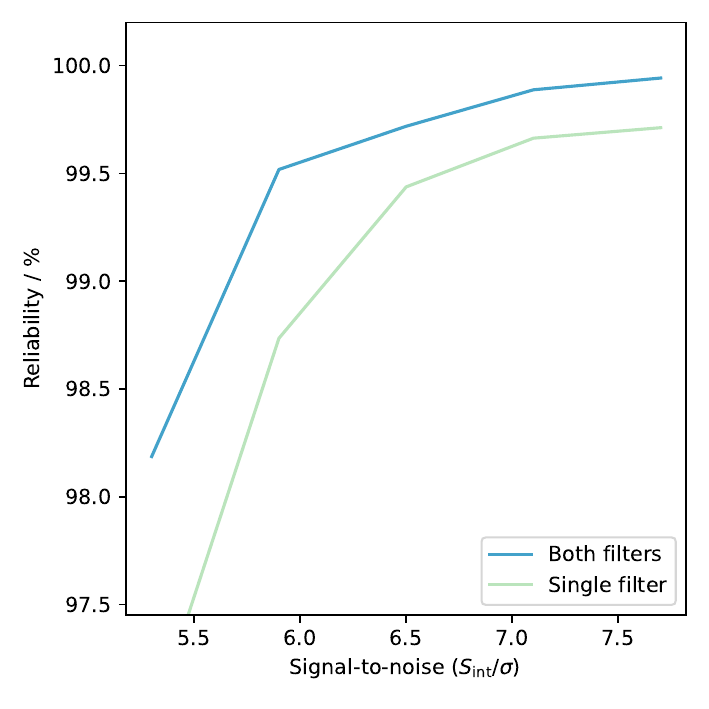}
    \end{subfigure}
    \caption{Estimates of the reliability of the catalogue as a function of the $S/N$. The left panel corresponds to $\ang{290} < l < \ang{44}$, while the right panel corresponds to $\ang{233} < l < \ang{324}$. In both panels, the lower green line is a conservative estimate before filtering out the sample near bright, positive sources. The upper blue curve is derived after these sources have been filtered out.}\label{fig:reliability}
\end{figure*}

To estimate the reliability of our source catalogue, we run \textsc{aegean} to detect negative sources in the mosaics using the same setting described in \sect~\ref{sec:catalogue}. Since real sources cannot have negative flux, these detections reflect noise and artefacts, allowing us to estimate the false detection rate. We then applied two filtering criteria. The first accounts for artefacts generated by very bright sources, which produce positive and negative features around themselves. These detections have been filtered out (positive and negative), accounting for the decreasing brightness of the artefacts as the distance from the source increases. This single-filter curve serves as a lower limit for the reliability of the catalogue. We then applied a second filter to the negative detections to discard all the elements within $2'$ of positive sources, likely generated by residual calibration errors. In total, we identified 156~negative sources within $\ang{290} < l < \ang{44}$ and 250 in $\ang{233} < l < \ang{324}$, resulting in an estimated catalogue reliabilities of $99.3\%$ and $99.4\%$, respectively. The reliability for each significant bin is illustrated in \figs~\ref{fig:reliability}.

We then applied the same approach implemented in the previous GLEAM \citep{Hurley2017} and GLEAM-X \citep{Hurley2022,Ross2024} surveys to quantify the completeness of the catalogue at 200\,MHz, which is the band used for source-finding, for the $\ang{290} < l < \ang{44}$ and $\ang{233} < l < \ang{324}$ regions. For each region, 15,000 point sources are simulated within the longitude range of the area covered using 21 realisations, each of which had a different flux density varying from $10^{-2}$ to $10^0$ Jy with an increment in the index of $0.1$. The simulated sources are chosen in a random position within $\ang{4} < |b| < \ang{11}$ and with a minimum separation of 5 arcminutes. Their major and minor axes are set to $a_{PSF}$ and $b_{PSF}$. \textsc{AeRes} is then used to inject these sources into the wide-band image. 

The same \textsc{Aegean} call used to detect sources in the 60\,MHz image for the source-finding procedure is applied here. The list of simulated sources recovered by the software is compared to their original sample, and their fraction is calculated to estimate the completeness. Simulated sources that lie close to a real source are considered successful only when the recovered position falls closer to the simulated source rather than the nearby real source.
%position of the simulated source instead of the bright source.

\begin{figure*}
    \centering
    \begin{subfigure}[b]{0.48\linewidth}
    \includegraphics[width=1\linewidth]{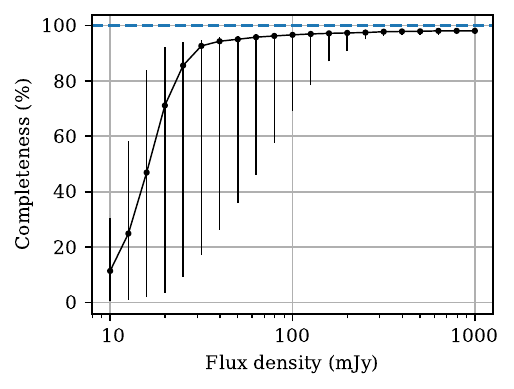}
    \end{subfigure}
    \begin{subfigure}[b]{0.48\linewidth}
    \includegraphics[width=1\linewidth]{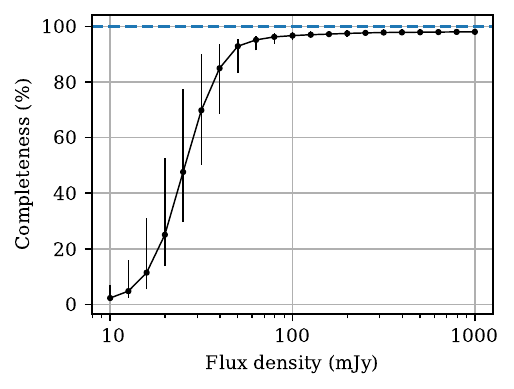}
    \end{subfigure}
    \caption{Estimates of the catalogue's completeness as a function of $S_{200 MHz}$. The left panel corresponds to $\ang{290} < l < \ang{44}$, while the right panel corresponds to $\ang{233} < l < \ang{324}$. The error bars are due to variations in completeness with dependence on $l$ because of the presence of bright sources.}\label{fig:completeness}
\end{figure*}

The completeness is estimated to be 50$\%$ at $\approx 25$\,mJy and increases to 90$\%$ at 50\,mJy in the $\ang{233} < l < \ang{324}$, while for the $\ang{290} < l < \ang{44}$ region it corresponds to 50$\%$ at $\approx 15$\,mJy and rising to 90$\%$ at 125\,mJy. The fraction of the simulated sources as a function of the $S_{200 MHz}$ is reported in \fig~\ref{fig:completeness} for both regions. The error bars reflect variations in completeness across the GP. In the longitude range $\ang{233} < l < \ang{324}$, the error bars are relatively small, indicating uniform completeness throughout the region. In contrast, the region $\ang{290} < l < \ang{44}$ shows larger error bars due to significant variations in completeness, ranging from a relatively quiet area near the Vela SNR and beyond to a much noisier region influenced by the presence of Centaurus~A.

\subsection{Comparison with GLEAM-GP} \label{sub:gleam}
\begin{figure}[h]
    \centering
    \includegraphics[width=1\linewidth]{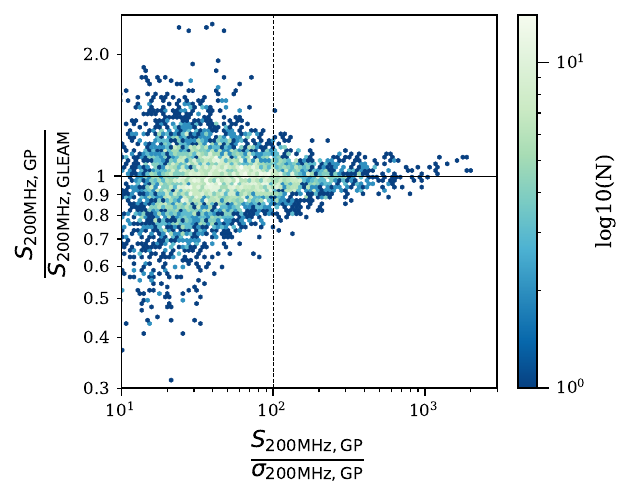}
    \caption{Ratio of the 200\,MHz integrated flux density for compact sources matched in the GP catalogue presented here and GLEAM-GP as a function of the $S/N$ in our sample. The vertical dashed line marks a $S/N$ of 100, corresponding to approximately 90$\%$ completeness in GLEAM. The horizontal solid line represents a ratio of 1, aligning with the overall trend. Colour indicates point density. Error bars are omitted for clarity but are calculated as the quadrature sum of measurement errors from both surveys.}\label{fig:gleam_fluxes}
\end{figure}

We use the GLEAM-GP data to calibrate flux density during data reduction (\sect~\ref{sec:catalogue}). We now compare the GLEAM-GP dataset to the GP catalogue presented here to examine any differences that might be in place between these two surveys. For this purpose, we considered only the available GLEAM-GP catalogue \citep[][]{Hurley2017} covering $\ang{345} < l < \ang{67}$, $\ang{180} < l < \ang{240}$ and galactic latitudes $\ang{1} < |b| < \ang{10}$. For the comparison, we select a compact subset of sources ($S_\mathrm{int}/S_\mathrm{peak} < 2$), that are well described by a power law (reduced-$\chi^2$ $< 1.93$, equivalent to a $99\%$ confidence level) and cross-matched within a radius of $15^{''}$. We additionally correct the GLEAM-GP data for the Eddington bias, which involves the flux density of faint sources being systematically biased high due to the noise \citep{Eddington1913}. Following the approach of \cite{Hurley2022}, we apply the correction from \eqn~4 of \cite{Hogg1998} to estimate the true flux density of GLEAM-GP sources at 200\,MHz to ultimately verify the GP flux density scale:
%, because there are more real faint sources that are biased high than bright ones that are fainter.
\begin{equation}
    S_{corr} = \dfrac{S_0}{2} \left( 1 + \sqrt{1 - \dfrac{4q + 4}{{S/N}^2}} \right)
\end{equation}

where q represents the number of sources to flux ratio in logarithmic scale and assumes a value of 1.54. The integrated flux density ratio is then calculated and shown in \fig~\ref{fig:gleam_fluxes} with the correction applied to the GLEAM flux densities. The line in black indicates the best trend for the fluxes, as expected around 1.

\begin{figure}[h]
    \centering
    \includegraphics[width=1\linewidth]{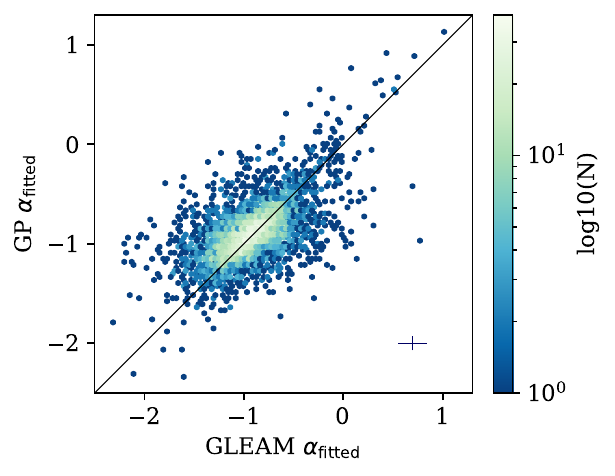}
    \caption{Spectral indices derived from a power law spectral model across the $20 \times 7.68$\,MHz narrow bands for compact sources matched in the GP catalogue presented here and GLEAM. The colour scale represents the density of points, and the average fitting error is shown as an error bar in the bottom right. The diagonal line indicates a 1:1 correspondence of $\alpha$.}\label{fig:gleam_alpha}
\end{figure}

We also compare spectral indices derived from a power law model between our GP catalogue and the published GLEAM-GP (after excluding all the sources with $\alpha$ set to 0 because we cannot find an optimal fit). The trend is shown in \fig~\ref{fig:gleam_alpha} and has the expected behaviour. In the bottom right corner of the illustration, we include the errors for both datasets: our GP catalogue shows significantly smaller errors due to the increased $S/N$ in this sky area, thanks to the introduction of GLEAM-X observations. We are then confident in the spectral fitting for the given sources. 

\begin{figure*}[h]
    \centering
    \includegraphics[width=1\linewidth]{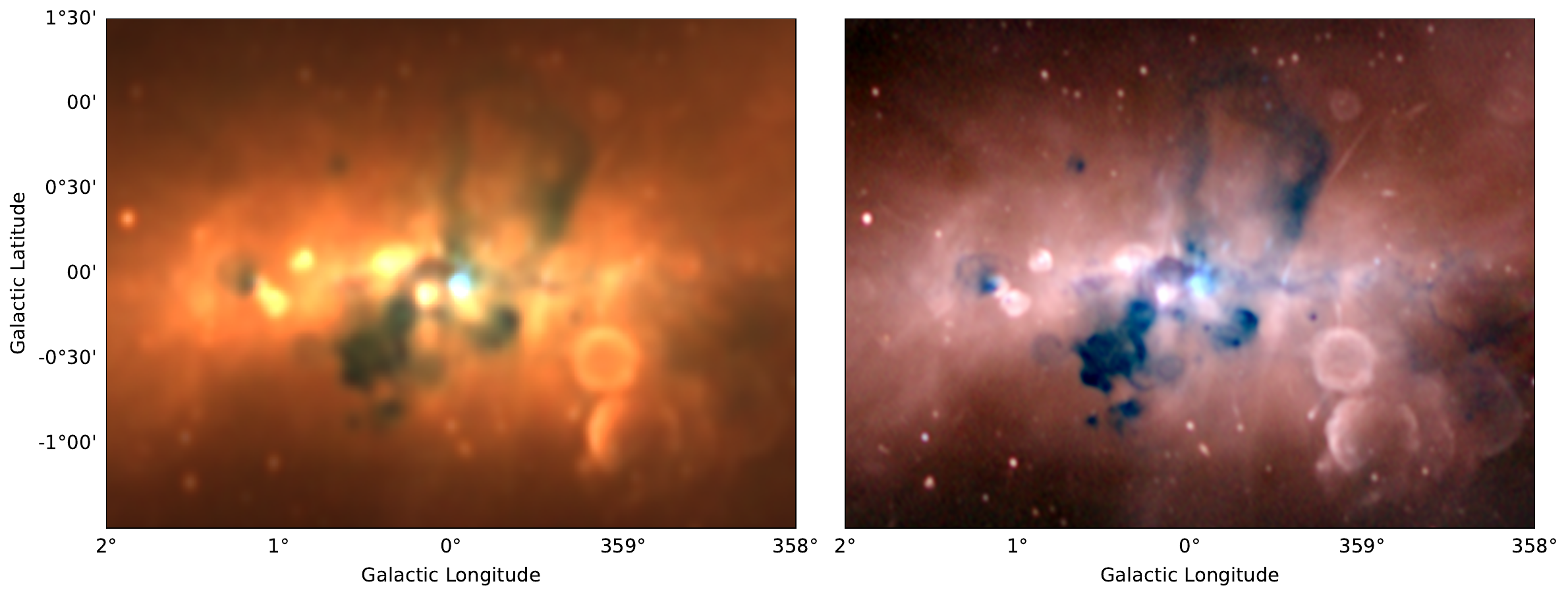}
    \caption{RGB cubes of the Galactic Centre as seen by GLEAM \citep[left panel; see ][ for a detailed analysis]{Hurley2017,Hurley2024} and by this data release (right panel). The cubes have been formed of the 72--103\,MHz (R), 103--134\,MHz (G), and 139--170\,MHz (B) data. The colour ranges used are -0.5--20~\Jypbm and -0.12--5~\Jypbm for the two panels, respectively.}\label{fig:gleam_centre}
\end{figure*}

A detailed comparison of the Galactic centre is presented in \fig~\ref{fig:gleam_centre}, highlighting the improved image quality relative to the GLEAM survey \citep{Hurley2019c}. Thanks to the inclusion of GLEAM-X observations, the new image shows a significant noise reduction, resulting in a cleaner background and a more distinct representation of structural features, offering a more precise view of its intricate components.

\section{Science applications}\label{sec:gp}
\subsection{Supernova remnants}
The frequency coverage within 72--231\,MHz combined with the sky area covered by this data release will likely be productive for studies of SNRs \citep[see ][for a review]{Dubner2015} in our Galaxy. SNRs go through four stages during their lifetime, where they exhibit different spectral behaviours according to the particle acceleration mechanisms that are taking place, how they interact with the surrounding medium, and finally, the amount of energy they are losing. Along the GP, we expect to observe $\approx$two~thousand SNRs \citep{Frail1994}. However, the actual number of objects we have been able to classify is around 300 \citep{Green2024}, less than half compared to the theoretical estimate. The lack of detections is likely caused by the limitations of previous radio surveys in terms of sky coverage and sensitivity to large scales. MWA has proven to be highly effective in detecting SNRs \citep{Hurley2019b,Mantovanini2025} due to its sensitivity to low-frequency radio emission, and thanks to the broad range of angular scales that we could recover ($45'' - \ang{15}$), we are confident this data release will help find more faint and old elements of the SNR population, helping to fill the current gap. 

\begin{figure}
    \centering
    \includegraphics[width=1\linewidth]{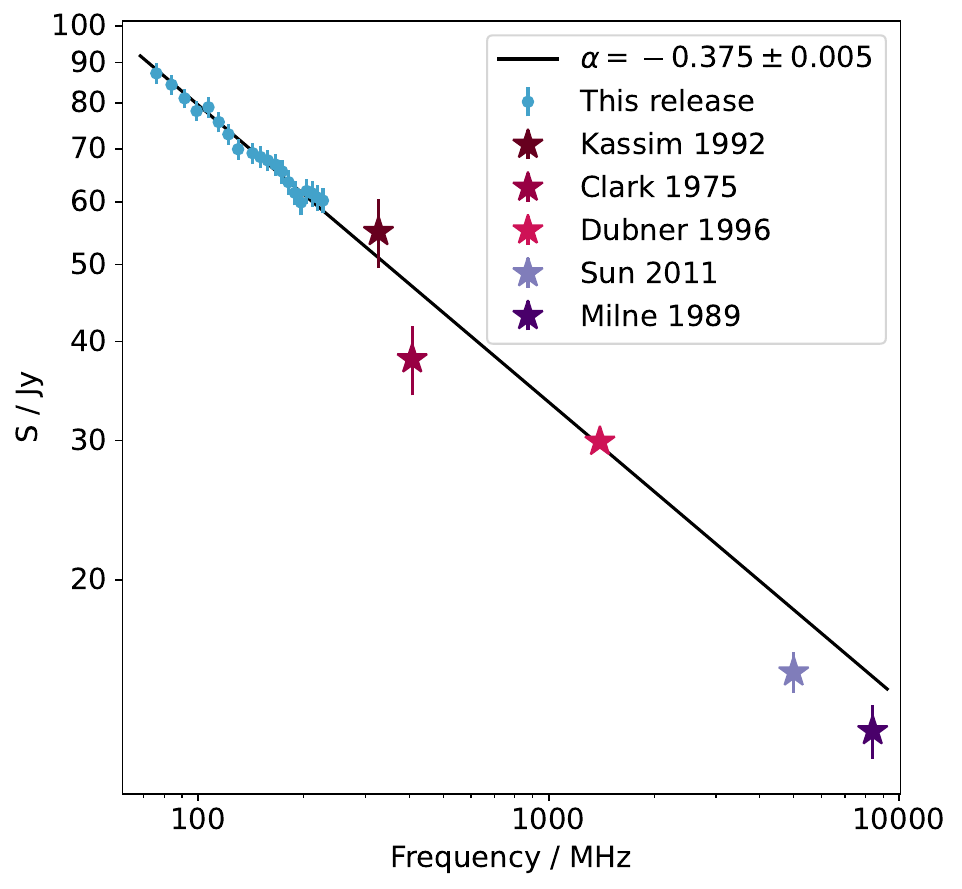}
    \caption{Radio continuum spectrum for G$18.8+0.3$ (Kes67) in logarithmic scale. The blue circles indicate the narrow band's flux density measurements presented in this work between 72--231\,MHz, while in other colours, the published measurements reported by \cite{Green2024} in the SNR catalogue from work by: \cite{Kassim1992} at 327\,MHz, \cite{Clark1975} at 408\,MHz, \cite{Dubner1996} at 1400\,MHz, \cite{Sun2011} at 5000\,MHz, and \cite{Milne1989} at 8400\,MHz. Where errors were not specified in the original papers, we assumed them to be $10\%$ of the flux density. The solid line represents the best-fitting curve.}\label{fig:snr}
\end{figure}

Furthermore, the sensitivity of our GP image will allow us to confidently measure the flux density of sources with an angular size up to a few degrees over a wide frequency band. It will enable detailed spectral fitting, essential in confirming the nature of the sources. We provide an example of the well-known G$18.8+0.3$ (also known as Kes67) spectra in \fig~\ref{fig:snr}. In addition, as highlighted by \cite{Castelletti2021}, these lower frequencies can provide insights into the physical properties of the observed regions, particularly for possible free-free absorption occurring inside or outside the remnant shell, which can be respectively caused by the presence of unshocked ejecta \citep[see ][ for a similar investigation]{Arias2018} or interaction with the surrounding environment (such as molecular clouds). In this regime, SNRs are better described by a power law with a turnover at frequencies $< 100$\,MHz \citep[as first stated by ][]{Dulk1975}.

On the other hand, the spectra of SNRs at radio frequencies are connected to gamma-ray emissions, particularly in the context of SNRs as potential sources of Galactic cosmic rays (CR) at very high energies \citep[up to PeV; further details in ][]{Cristofari2020}. Therefore, the radio spectrum can constrain the efficiency and nature of electron acceleration within SNR shocks. Since the processes driving proton acceleration are expected to be similar, radio studies are also relevant to investigations into the origin of CRs \citep[as outlined in ][ and references therein]{Dubner2015}. It is then worth looking for possible associations between gamma rays and SNRs \citep{Maxted2019}. The GP image could be crucial in identifying faint radio structures previously gone unnoticed, with which a gamma-ray source may be associated.

\subsection{\textsc{hii} regions}
\textsc{Hii} regions are mostly formed by the ionisation of the medium due to the ultraviolet radiation emitted by a newly born star \citep{Condon2016}; as such, they represent good tracers for star formation regions in the galaxy. These sources have been mostly identified at infrared (IR) wavelength; the Widefield Infrared Survey Explorer \citep[AllWISE;][]{Wright2010,Mainzer2011} have been extensively used, where all sources show essentially the same morphology: at $\simeq 22$\;$\mu$m, the main contribution to the emission is caused by the radiation of hot dust, and it is surrounded by a ring at $\simeq 12$\;$\mu$m caused by Polycyclic Aromatic Hydrocarbon molecules excited by UV radiation from the nearby stars \citep[][]{Watson2008}. The ionised gas is visible at radio frequencies, which are then used to confirm the identification of a \textsc{hii} region. 

\begin{figure*}
    \centering
    \includegraphics[width=1\linewidth]{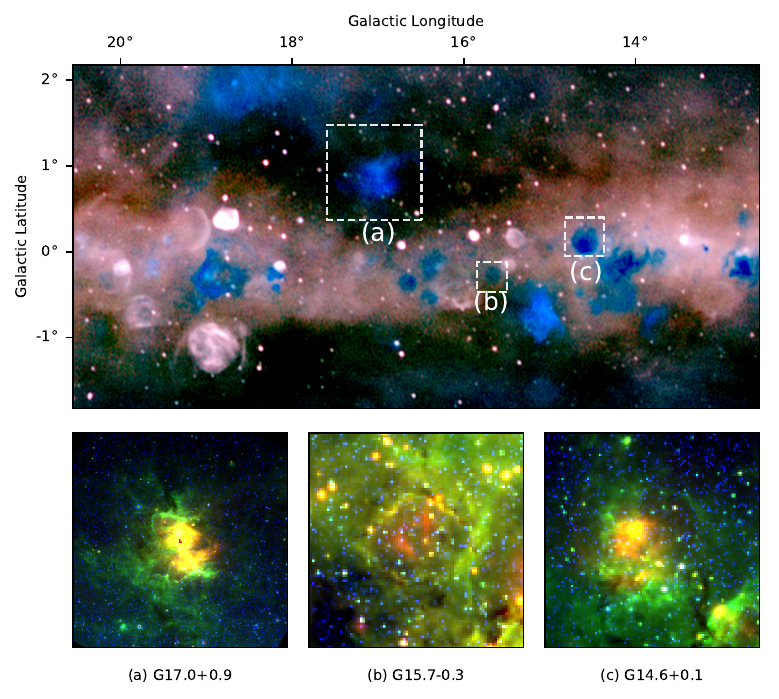}
    \caption{Section of $\ang{8}$ in longitude of the southern GP. The top panel shows an RGB cube formed of processed data from this release: 72--103\,MHz (R), 103--134\,MHz (G), and 139--170\,MHz (B). White squares indicate three examples of known \textsc{hii} regions in the area: G17.0$+$0.9 (a), G15.7$-$0.3 (b), and G14.6$+$0.1 (c). The bottom row shows corresponding objects in AllWISE bands, with the RGB channels assigned to $22$\;$\mu$m (R), $12$\;$\mu$m (G), and $3.4$\;$\mu$m (B).}\label{fig:hii}
\end{figure*}

\textsc{hii} regions are typically dominated by thermal free-free emission with a spectral index between $-0.2 < \alpha < +2$. They become optically thick at low radio frequencies (below 100\,MHz), which means that the diffuse non-thermal emission of the background is absorbed. This effect is visible in our observations. We constructed an RGB cube composed of the 72--103\,MHz (R), 103--134\,MHz (G), and 139--170\,MHz (B) images, as shown in the top panel of \fig~\ref{fig:hii}. \textsc{hii} regions are immediately recognised by a distinctive blue colour in opposition to the non-thermal emission, which appears red due to its negative spectral index. The boxes represent three known objects, and their appearance at IR wavelengths is shown in the smaller panels of the second row. The different colours will be crucial in disentangling the emission's origin, particularly in SNR identification, for which these thermal sources are often a source of confusion \citep{Helfand2006,Hurley2019a}. 

\cite{Hindson2016} leveraged their distinctive blue colour to compile a comprehensive catalogue using GLEAM data over the longitude range $\ang{340} < l < \ang{260}$. The final result has been implemented \citep[see ][]{Polderman2019,Polderman2020} to measure the synchrotron radiation emitted along the line of sight (the so-called synchrotron emissivity $\epsilon$) in front and behind absorbed \textsc{hii} regions. A broader study by \cite{Su2018} mapped the synchrotron emissivity throughout the Galaxy using 152~\textsc{hii} regions. By deriving optical depths, the authors achieved more precise estimates of how much radiation each object absorbs, enabling more reliable measurements of $\epsilon$. The synchrotron emissivity is directly related to the density of CR electrons and the strength of the GMF component perpendicular to the line of sight. The approach then manages to constrain the CR population and the structure of the GMF across the plane. However, it is challenging to disentangle the contribution of each component, and precise models are necessary to address this degeneracy and understand their role in our galaxy's constitution. The increased resolution and sensitivity of the GP survey presented here may be essential in detecting more absorption features and enabling emissivity calculation over a wider range of lines of sight.

\subsection{Planetary Nebulae}
Planetary nebulae (PNe) mark the final evolutionary stages of low to intermediate mass ($\sim$1--8\,\(\textup{M}_\odot\)) stars. The large stellar winds from these stars, while in the asymptotic giant branch (AGB) phase, result in significant mass loss. Radio continuum observations of PNe reveal the thermal free-free emission from ionised outflows, and are not limited by the dust extinction in the Milky Way like optical observations \citep{2009A&A...498..463P,2015MNRAS.451.3228C}. PNe typically become optically thin with a spectral peak at frequencies dependent on the age and angular size of the PNe. Characterising radio continuum SEDs, particularly over the spectral peak and optically thick region, can help to characterise the properties of PNe, resolving long-standing tensions between different geometries, density and temperature profiles and velocities. The region included in this data release provides crucial insights into the optically thick frequencies for several known or candidate PNe. For example, NGC~3918 (PN G294.6+04.7), a well known and studied bright PN, is modelled with a power-law spectral model with a spectral index of $\alpha=1.2$, consistent with previous radio observations \citep{2015MNRAS.451.3228C}. 

\subsection{Pulsars}
Continuum radio images, while less traditional, are a novel way of detecting low-frequency pulsars that may be undetectable in high time resolution searches due to scattering and dispersion, and can provide additional information on the sources, such as accurate flux density measurements. \cite{Sett2024} provides an example of this approach at $\leq 300$~MHz, detecting 83 known pulsars, 14 of which were the first reported low-frequency detections, previously missed by periodic searches because located at high Dispersion Measure or highly scattered. A simple power law with a very steep index \citep[on average $-1.6$ as illustrated by ][]{Bates2013} is often used to describe pulsars; in $10\%$ of the known pulsars though, it is observed a turnover at low frequencies (below 400\,MHz) that may be caused by synchrotron self-absorption or thermal absorption by gas present along the same line of sight of the source \citep[as described in ][and references therein]{Swainston2021,Swainston2022}. Thanks to the frequency range covered in this work, the survey of the GP we are presenting here can be crucial in bringing out the presence of a turnover in the spectra of known pulsars. \cite{Murphy2017} performed similar work on a limited sample of pulsars visible by GLEAM at 200\,MHz. \citep{Mantovanini2025b} show that the reduced confusion noise enabled by introducing GLEAM-X observations increases the percentage of the population visible at these frequencies to examine the overall population better, and provides new pulsar candidates ideal for follow-up pulsation searches at higher frequencies. Furthermore, cross-matching steep-spectrum sources in our catalogue with unassociated gamma-ray sources may reveal additional pulsar candidates \citep{Frail2018} worth following up in future studies, and an analysis of their eclipsing variability could help identify them as eclipsing binary millisecond pulsars (MSPs; Petrou et al., in prep.).

\section{Final remarks}\label{sec:conclusion}
This work provides $\approx 3800$\,deg$^2$ of the southern GP as part of the GLEAM-X survey. The $26 \times 2$ mosaics cover an area of $\ang{233} < l < \ang{44}$ within latitudes of $\abs{b} < \ang{11}$ generated by a combination of 1,898 2-minute observations corresponding to the $57\%$ of the total number of processed data. $53\%$ of the successful observations are from GLEAM-X, and $47\%$ from GLEAM. Along with the mosaics, we also provide a source catalogue composed of 98207 elements, for which 92787 were successfully fitted either by a simple or curved power law.

\begin{acknowledgement}
N.H.-W. is the recipient of an Australian Research Council Future Fellowship (project number FT190100231). 

This scientific work uses data obtained from Inyarrimanha Ilgari Bundara, the CSIRO Murchison Radio-astronomy Observatory. Support for the operation of the MWA is provided by the Australian Government (NCRIS), under a contract to Curtin University administered by Astronomy Australia Limited.
We acknowledge the Pawsey Supercomputing Centre which is supported by the Western Australian and Australian Governments.
\end{acknowledgement}

\printendnotes
%\printbibliography
\bibliography{Biblio}

\begin{appendix}
\section{Ionospheric activity}\label{app:iono}
As mentioned in \sect~\ref{sub:astrometric}, to assess the impact of ionospheric conditions on data quality, we analysed the distribution of pipeline failures across the five 30\,MHz bands used in this survey. The results show that the lower-frequency band (72–103 MHz) is significantly more susceptible to failure, particularly during imaging. These increased losses are primarily attributed to enhanced RFI contamination and ionospheric turbulence.

\begin{figure}[th!]
    \centering
    \includegraphics[width=1\linewidth]{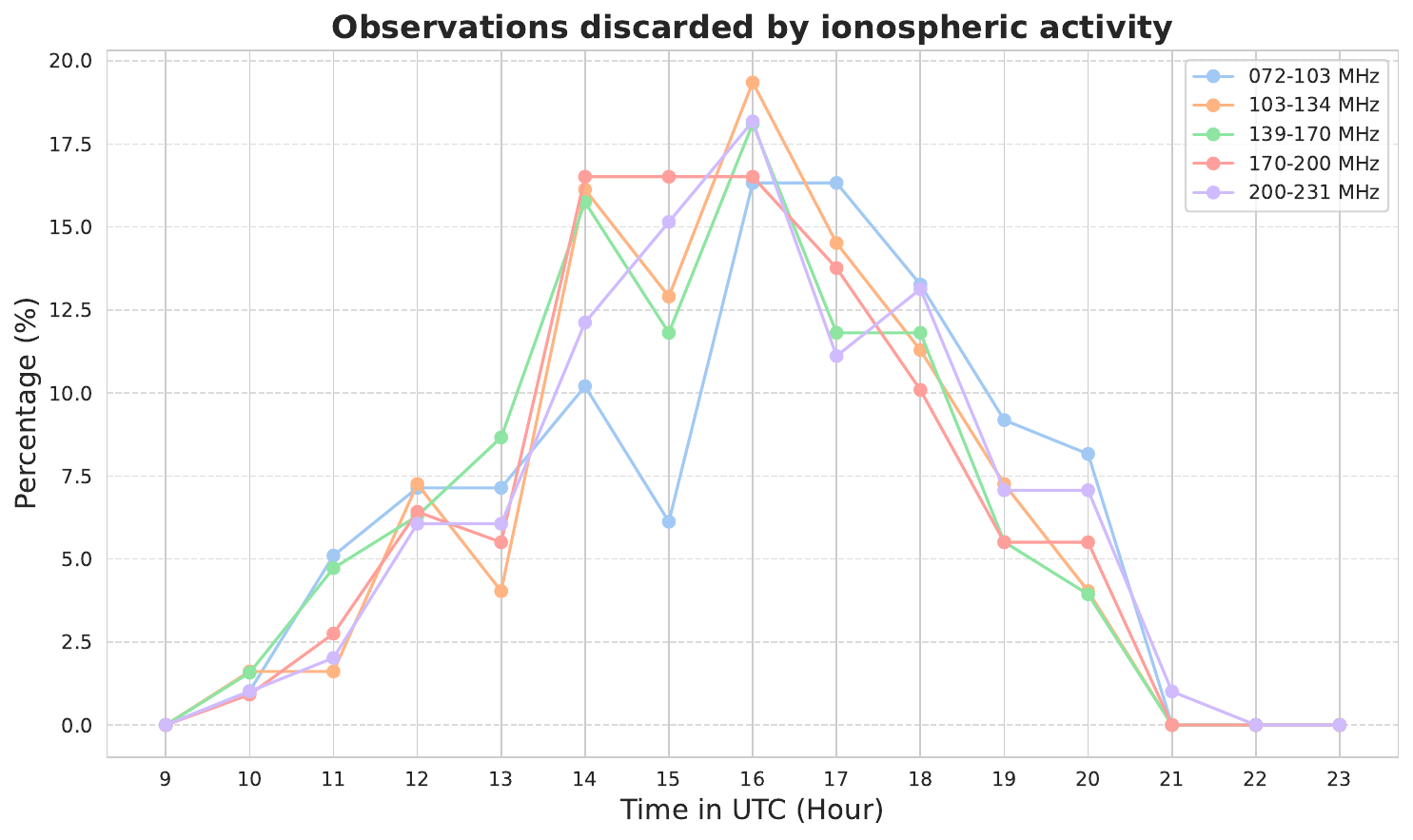}
    \caption{Distribution of the observations discarded across the five 30\,MHz frequency channels due to strong ionospheric activity.}\label{fig:iono}
\end{figure}

To quantify the impact of ionospheric activity, \fig~\ref{fig:iono} presents the fraction of observations discarded as a function of time, clearly illustrating periods of increased data loss across all five frequency channels, probably caused by ionospheric disturbances.

\section{Noise levels} \label{app:noise}
\begin{figure}[th!]
    \centering
    \includegraphics[width=1\linewidth]{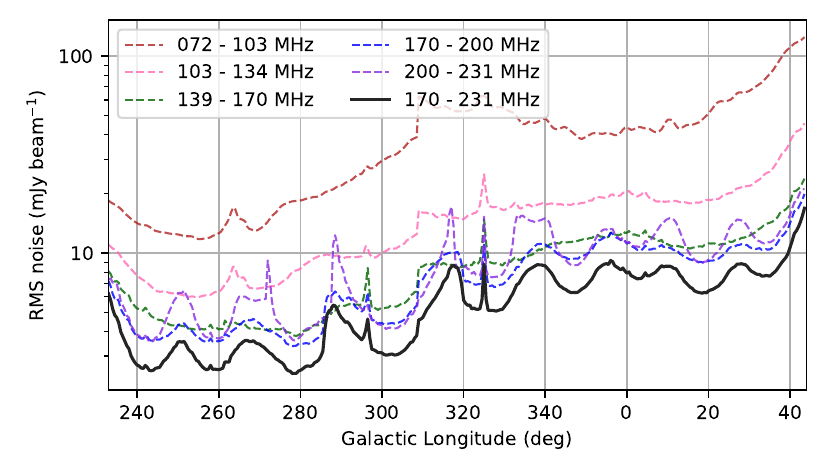} \\
    \includegraphics[width=1\linewidth]{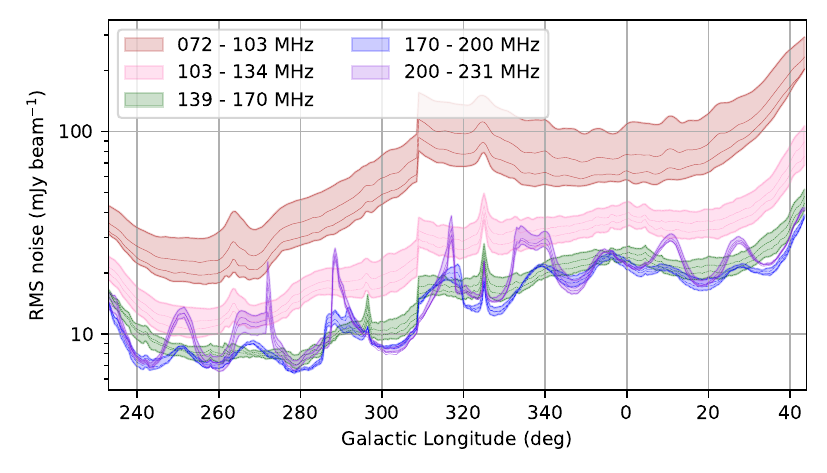}
    \caption{Variation of RMS noise across the GLEAM-X: GP longitude range. Top panel: RMS trends for the five $30$\,MHz bands (dashed lines) and the wideband image. Bottom panel: RMS trends for the $20 \times 7.68$\,MHz sub-bands. Shaded regions indicate the frequency ranges corresponding to the four bands that make up the broader images.}\label{fig:noises}
\end{figure}

The RMS noise across the GLEAM-X: GP region is shown in \fig~\ref{fig:noises}. Noise levels generally increase toward the Galactic centre and at northern hemisphere longitudes. Peaks observed across the five frequency bands are primarily caused by the presence of sources; for example, the peak near longitude $325^{\circ}$ corresponds to a cluster. An exception is the $200 - 231$\,MHz band, where the peaks are also influenced by reduced mosaic quality, which results from a smaller field of view at higher frequencies, leading to less coverage during the mosaicking process compared to lower channels. However, this trend does not significantly affect the wideband image, which maintains values $< 10$~\mJypbm across most of the longitude range.

\section{$(u,v)$ coverage} \label{app:uvcov}
\begin{figure}[th!]
    \centering
    \includegraphics[width=1\linewidth]{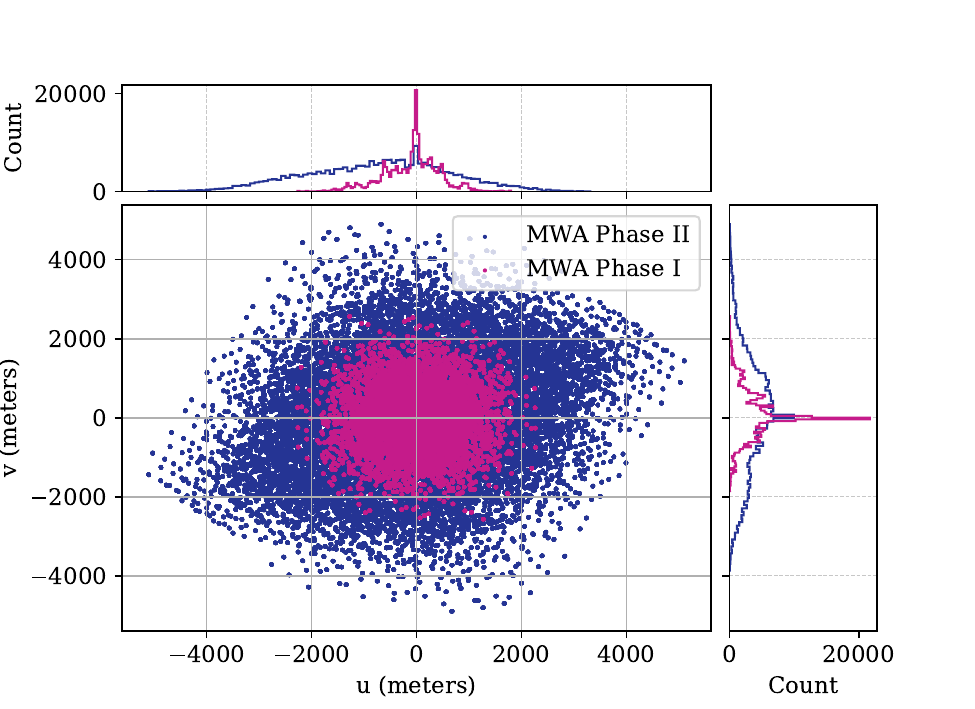}
    \caption{$(u,v)$-coverage for the MWA Phase~\textsc{i} (blue) and Phase~\textsc{ii} (red) configurations.} \label{fig:uvcov}
\end{figure}

A visualisation of the instrument Phase~\textsc{i} and Phase~\textsc{ii} $(u,v)$-coverage is reported in \fig~\ref{fig:uvcov}. The dense core of the MWA Phase~\textsc{i} configuration is key for detecting the larger structures of the Galaxy. In the ``extended'' Phase~\textsc{ii} configuration, the abundance of short baselines has been reduced, and hence dropping sharply the sensitivity above $\approx 20^{'}$; the longest baseline, instead, has been doubled in order to gain sensitivity to the finest scales. 

\section{Observations employed}\label{app:obsids}
The successful 1898 observations used in this work are compiled in a CSV table. For each observation, we provide the central frequency, start time, coordinates (in both FK5 and Galactic systems), and pointing information (elevation and azimuth). Observations are grouped by frequency channel and ordered sequentially in time. The initial four entries are presented in \tab~\ref{tab:obsids}. 

\onecolumn
\begin{center}
\topcaption{Example of a summary table comprising the list of GLEAM and GLEAM-X observations used to create the final mosaic of the GP for channel 72--103\,MHz. A table is generated for all five frequency channels. The table includes the observation ID, the central frequency, the time the observation started, the Azimuth (Az) and Elevation (El) of the pointing, the RA and Dec coordinates in degrees, and the l and b coordinates in degrees.}\label{tab:obsids}
\tablefirsthead{\toprule Observation ID & Freq. (MHz) & Start Time (UTC) & Az($^{\circ}$) & El($^{\circ}$) & RA($^{\circ}$) & Dec($^{\circ}$) & l($^{\circ}$) & b($^{\circ}$)\\ \midrule}
\tablelasttail{\bottomrule}
\begin{supertabular}{lllllllll}
1079096624 & 87.68 & 2014-03-17 13:43:28 & 180.0 & 61.7 & 137.5 & -54.9 & 274.6 & -4.8 \\ 
1102969536 & 87.68 & 2014-12-18 20:25:20 & 194.0 & 60.7 & 138.4 & -54.6 & 274.7 & -4.1 \\ 
1204209616 & 87.68 & 2018-03-04 14:39:58 & 180.0 & 61.7 & 138.8 & -54.9 & 275.1 & -4.2 \\
1204982096 & 87.68 & 2018-03-13 13:14:38 & 166.0 & 60.7 & 138.1 & -54.6 & 274.6 & -4.2 \\
\end{supertabular}
\end{center}
\end{appendix} 

\end{document}